\documentclass[journal]{IEEEtran}
\usepackage{cite}
\usepackage{algorithm}
\usepackage{algorithmic}
\usepackage{breqn}
\usepackage{colortbl,booktabs}
\usepackage{tabularx}
\usepackage{enumitem}
\usepackage{graphicx}
\usepackage{float}
\usepackage{amsmath, amsthm, amssymb, amsfonts, mathtools}
\usepackage{bm}
\usepackage{textcomp}
\usepackage{float}
\usepackage{color}
\usepackage{todonotes}
\usepackage{gensymb}

\usepackage{multirow} 
\usepackage{array}
\usepackage{stfloats}

\usepackage{cite}

\normalsize

\begin{document}

\title{In-memory Multi-valued Associative Processor}

\author{Mira Hout, Mohammed E. Fouda, Rouwaida Kanj, and Ahmed M. Eltawil
\thanks{M. Hout and R. Kanj are with ECE Dept., American University of Beirut, Lebanon, 1107 202}
\thanks{M. Fouda is with Center for Embedded \& Cyber-physical Systems, University of California-Irvine, Irvine, CA, USA 92697-2625, with Engineering Mathematics and Physics Dept., Faculty of Engineering, Fayoum University, Egypt and with Nanoelectronics Integrated Systems Center (NISC), Nile University, Giza, Egypt. (Email:foudam@uci.edu) }
\thanks{A. Eltawil is with King Abdullah University of Science and Technology (KAUST), Thuwal 23955, Saudi Arabia and with Center for Embedded \& Cyber-physical Systems, University of California-Irvine, Irvine, CA, USA 92697-2625}
\thanks{Manuscript received xxx, xxx; revised xxxx, xxx.}}

\markboth{}%
{Hout \MakeLowercase{\textit{et al.}}: Multi-valued Associative Processor}

\maketitle

\begin{abstract}

In-memory associative processor architectures are offered as a great candidate to overcome memory-wall bottleneck and to enable vector/parallel arithmetic operations. In this paper, we extend the functionality of the associative processor to multi-valued arithmetic. To allow for in-memory compute implementation of arithmetic or logic functions, we propose a structured methodology enabling the automatic generation of the corresponding look-up tables (LUTs). We propose two approaches to build the LUTs: a first approach that formalizes the intuition behind LUT pass ordering and a more optimized approach that reduces the number of required write cycles. To demonstrate these methodologies, we present a novel ternary associative processor (TAP) architecture that is employed to implement efficient ternary vector in-place addition. A SPICE-MATLAB co-simulator is implemented to test the functionality of the TAP and to evaluate the performance of the proposed AP ternary in-place adder implementations in terms of energy, delay, and area. Results show that compared to the binary AP adder, the ternary AP adder results in a 12.25\% and 6.2\% reduction in energy and area, respectively. The ternary AP also demonstrates a 52.64\% reduction in energy and a delay that is up to 9.5x smaller when compared to a state-of-art ternary carry-lookahead adder.

\end{abstract}

\begin{IEEEkeywords}
Associative Processor, Multi-valued Logic, Ternary, Adder, In-memory Computing, Compute-in-Memory.
\end{IEEEkeywords}

\IEEEpeerreviewmaketitle
\section{Introduction}
%
%
%
%

\IEEEPARstart{M}{any} applications in Machine Learning, big data analysis, search engines, and network routing require massive parallelism. As the amount of data to be analyzed continues to grow, power dissipation and data transfer between memory and processing units have limited the scalability of parallel architectures \cite{white_paper}.

This has led researchers to consider Associative Processors (APs) as in-memory platforms for carrying-out massively parallel computations inside the memory without the need to move the data \cite{AP}. The ability of APs to unify data processing and data storage has drastically decreased compute energy and latency cost. An AP constitutes of an array of Content Addressable Memories (CAMs) which are storage devices that allow concurrent access to all data rows. 
A CAM searches for a stored word based on an inputted search key. The AP improves its functionality by allowing parallel writing into masked bits of the matching CAM rows \cite{CAM}.

While binary CAMs perform exact-match searches for `0' and `1' bits, a more powerful ternary CAM (TCAM) can search a third “don’t care” value, allowing for very flexible pattern matching between search keys and stored values \cite{CAM}. Considering its promising applications potential, different implementations for the TCAM have been proposed including SRAM-based TCAM \cite{sram_tcam, sram_tcam2}. However, they were not widely adopted due to their low density and nonvolatility \cite{binary_AP}. Nowadays, emerging devices such as resistive memories have relaxed these constraints, leading to the revival of the TCAM-based AP approach in the research community \cite{ReRAM_AP}. New TCAM implementations based on resistive random access memory (ReRAM) have been proposed \cite{MCAM, MCAM2} to reduce power and improve area density in comparison to the conventional complementary metal-oxide semiconductor (CMOS) solutions. 

Memristive-based TCAM (MTCAM) designs have been proposed to build 1D and 2D in-memory computing architectures based on AP \cite{yavits2015resistive, binary_AP}. The AP architecture presented in \cite{binary_AP} was designed to perform in-memory compute in the context of the binary full adder. It relies on a look-up table (LUT)-based 1-bit addition that employs four compare and write operations applied in parallel to the different rows, resulting in significant runtime savings. 

Recently, ternary logic has gained interest among the circuit design community for its ability to increase power efficiency and reduce the complexity of arithmetic circuit designs. However, the implementation of ternary logic circuits requires the use of devices with multiple threshold voltages which is difficult to accomplish with the current CMOS technology with reasonable devices' area and latency \cite{CNTFET,mohammaden2021cntfet}. Therefore, recent studies have shed light on alternative devices for the design of ternary arithmetic circuits such as carbon nanotube field-effect transistors (CNTFETs) \cite{CNTFET,CNTFET2,mohammaden2021cntfet} and memristors \cite{nancy,other_adders}. 

In order to carry-out ternary addition, different approaches were adopted in the literature. In \cite{CNTFET}, authors included a ternary-to-binary decoder stage for the inputs, and the addition was performed using binary logic gates before converting back the outputs to ternary logic. In \cite{CNTFET2}, the authors expressed the arithmetic function using a K-Map, and they used the obtained equations to determine the logic gates needed for the CNTFET-based implementation. In \cite{nancy}, the authors custom-designed the desired arithmetic function and implemented it using ternary logic gates composed of both memristors and CNTFETs.

In this paper, we propose a scalable CAM cell design and methodology for purposes of multi-valued logic AP applications. To enable the implementation of in-memory compute operations, we propose two novel algorithms that guide the automatic generation of the LUT for in-place multi-valued arithmetic or logic functions. The first relies on a depth-first search exploration of the state diagram obtained from the function's truth table, and the second capitalizes on common outputs to reduce the number of required write cycles.
The proposed methodology is universal and can be employed for different logic or arithmetic functions such as NOR, XOR, AND, multiplication, addition and subtraction. To illustrate, we present a novel implementation of a ternary AP (TAP) architecture based on a novel quaternary CAM (QCAM) design. We demonstrate the proposed design in the context of a LUT-based ternary full adder application.
Specifically, our contributions are as follows: 
\begin{enumerate}
 \item We propose a ``$n$T$n$R'' CAM cell for multi-valued AP (MvAP) arithmetic and logic operations. To exemplify our design, we present a TAP architecture using a ``3T3R'' QCAM cell as a building block. 
 \item We propose a scalable cycle-free state diagram mapping of the multi-valued arithmetic or logic function's truth table. The state diagram forms the core data structure to implement the proposed algorithms. These algorithms build on the states' connectivity along with other relevant attributes to structurally traverse the state diagram for a systematic generation of the LUTs.
 \item A first approach that relies on depth-first search (DFS) parsing of the state diagram to determine the order of the passes. 
 \item A second optimized approach that exploits common write actions to reduce the number of required write cycles. It relies on breadth-first search (BFS) parsing and a grouping heuristic for the different state diagram nodes.
 \item We test the algorithms for implementing a LUT-based ternary full adder (TFA) relying on a TAP architecture. We evaluate the energy, delay and area efficiency of the ternary adder implementations using the first and second approaches and compare them against each other as well as to the AP binary adder and other ternary full adder implementations.
\end{enumerate}

The rest of this paper is organized as follows. 
Section \ref{sec:MvAP} proposes the multi-valued AP architecture and discusses the CAM implementation and operation. Then, an illustrative example of ternary AP is discussed in Section \ref{sec:TAP}. The extended functional operations from binary AP is proposed utilizing a novel state diagram interpretation in Section \ref{sec:AP_Operation}, and an optimized version is introduced in Section \ref{sec:Optimized_AP}. Experimentation results and analysis are performed on a novel ternary AP full adder implementation in Section \ref{sec:results}. Finally, the conclusion of this work is given.





\section{Proposed MvAP Architecture}
\label{sec:MvAP}

Traditionally, digital arithmetic computation is performed using two-valued logic: `0' or `1'. However, in modern digital design, researchers are increasingly looking into multi-valued logic (MVL) as a way to replace the classical binary characterization of variables. MVL, notably ternary logic, constitutes a promising alternative to the traditional binary logic \cite{MVL_ternary} as it provides multiple advantages such as reduced interconnects, smaller chip area, higher operating speeds and less power consumption \cite{ternary_advantages}. The voltage levels for multi-valued logic of radix-$n$ are $n$ levels spanning from $0$ to $V_{DD}$. Hence, the $i^{th}$ logic value is realized with $i*V_{DD}/(n-1)$ where $i\in [0,n-1]$. For instance, ternary logic system uses $\{0,\,1,\, 2\}$ logic values with $\{0,\, V_{DD}/2,\, V_{DD}\}$ voltage levels, respectively. This representation is referred to as the unbalanced representation unlike the balanced which uses $\{-1,\, 0,\, 1\}$ logic values and is realized with $\{-V_{DD},\, 0,\, V_{DD}\}$ voltage levels \cite{ternary_systems}. In this paper, we focus on the unbalanced ternary logic system. 

\begin{figure*}[]
\centering
\includegraphics[width=1\linewidth]
{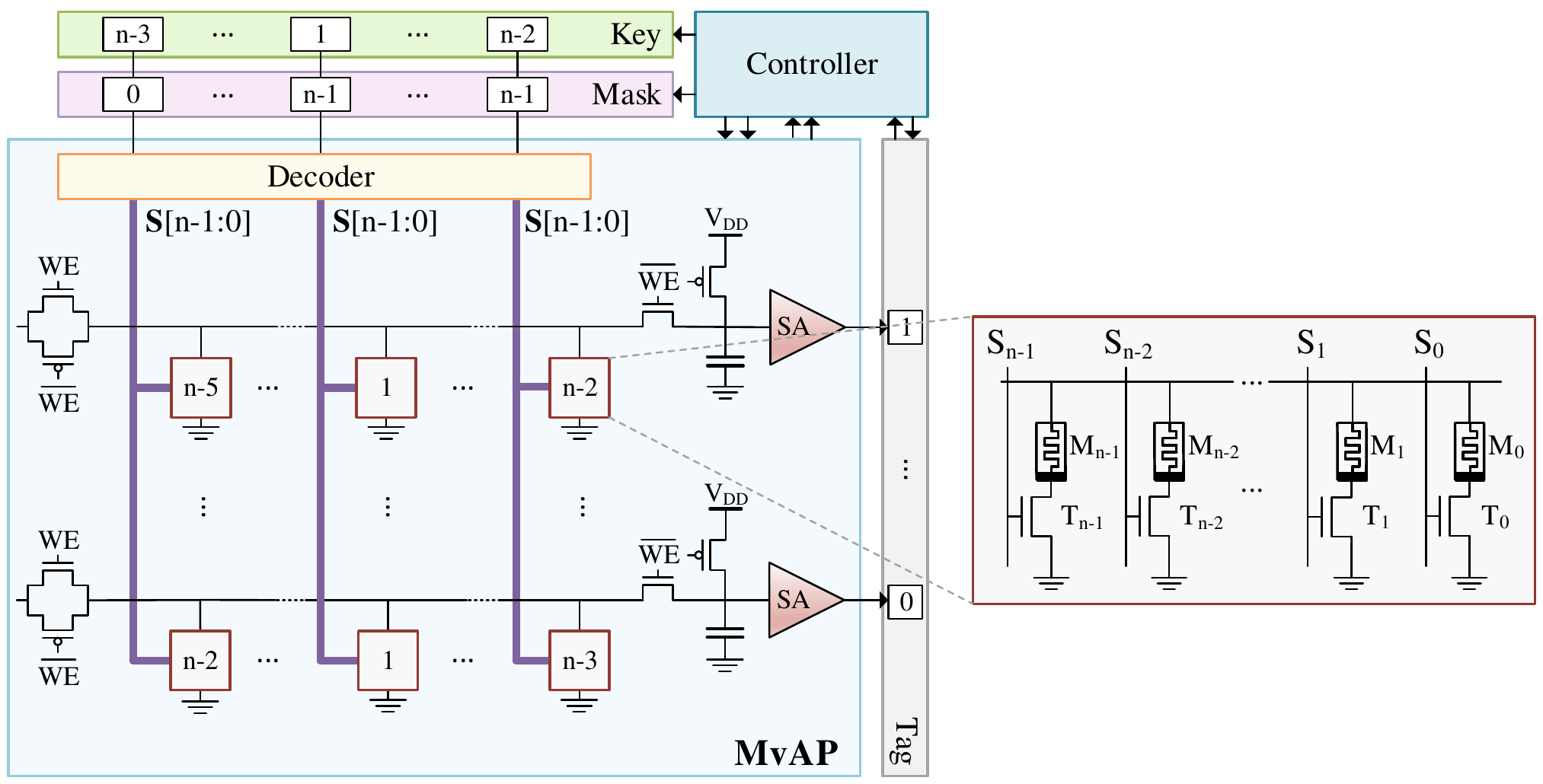}
\caption{An illustration of the MvAP architecture highlighting the proposed MvCAM cell.}
\label{fig:mvap}
\vspace{-0.1in}
\end{figure*}

Fig. \ref{fig:mvap} presents an illustration of the MvAP architecture comprised of a multi-valued CAM (MvCAM) array, a controller, a decoder and a set of Key, Mask and Tag registers. A MvCAM array consists of multiple rows containing MvCAM cells where $n$-valued digits (nits) are stored. The following sections present the proposed implementation of the different components.

\subsection{MvCAM Cell}
A compact design of a memristor-based MvCAM cell with $n$ transistors and $n$ memristors (``$n$T$n$R'') is presented as a natural extension to the ``2T2R'' TCAM cell designated for binary AP applications \cite{2T2R_fouda}. The proposed design is illustrated in the context of a multi-valued AP in Fig. \ref{fig:mvap}. The memristors in the cell function as storage elements whose states determine the stored nit value. The stored value is obtained by setting only one of the memristors to the low resistance state $R_{LRS}$ and maintaining the other $(n-1)$ memristors in the high resistance state $R_{HRS}$, as indicated in Table \ref{tab:general_stored}. Without loss of generality, the location of the single $R_{LRS}$ among the $(n-1)$ remaining $R_{HRS}$ memristors determines the logic state stored in the cell. Specifically, to store nit value $i$, memristor $M_i$ is the one set to $R_{LRS}$. A ``don't care'' state is represented by all memristors set to $R_{HRS}$.
To test the functionality of the ``$n$T$n$R'' cell, the matchline (ML) is initially precharged high.
The signal vector $(S_{n-1},\, S_{n-2}\, ...\, S_1,\, S_0)$ illustrated in Fig. \ref{fig:mvap} is then sent to check for a specific stored nit value in the cell. When searching for nit value $i$, signal $S_i$ is set to low while the other signals are set to high. The search outcome results in a match only when memristor $M_i$ is in the $R_{LRS}$ and the other memristors are in the $R_{HRS}$. Otherwise, the search outcome results in a mismatch. In the case of a match, the voltage of the ML discharges slowly and is hence preserved high, whereas in the case of a mismatch, the ML discharges quickly to ground.

\begin{table}[]
\centering
\caption{Mapping between the nit value stored in the MvCAM cell and the corresponding $n$ memristor states.}
\label{tab:general_stored}
\begin{tabular}{|c|ccccc|}
\hline
\multirow{2}{*}{\begin{tabular}[c]{@{}c@{}}Logic\\ value\end{tabular}} & \multicolumn{5}{c|}{Stored state} \\
 & $M_{n-1}$ & $M_{n-2}$ & $\cdots$ & $M_1$ & $M_0$ \\ \hline
x & H   & H   & $\cdots$ & H  & H  \\
0 & H   & H   & $\cdots$ & H  & L  \\
1 & H   & H   & $\cdots$ & L  & H  \\
$\vdots$ &$\vdots$ & $\vdots$ & $\ddots$ & $\vdots$ & $\vdots$\\
$n-2$ & H   & L   & $\cdots$ & H  & H  \\
$n-1$ & L   & H   & $\cdots$ & H  & H  \\ \hline
\end{tabular}
\vspace{-0.1in}
\end{table}

\subsection{Search Key $n$-ary Decoder}
The $n$-ary decoder allows mapping the key-mask pair to the signal vector $(S_{n-1},\, S_{n-2}\, ...\, S_1,\, S_0)$. Table \ref{tab:general_decoder} presents the truth table for the $n$-ary decoder. The inputs to the decoder are a binary mask and nit-valued key. In the signal outputs of the decoder, the place of the signal set to zero is equal to the search key value. Specifically, to search for logic value $j$, the signal $S_j$ is the one set to zero. It is worth noting that the decoder logic is inverting since the target signal is set to low whereas all other signals are set to high. When the key is masked, i.e., the mask bit is a zero, all decoded signals are set to zero. One simple and generic way to implement such decoder is with simple successive approximation ADC with modified operation. But, in the case of the ternary decoder, it will be shown in the next section that it can be realized with some ternary logic circuits.

\begin{table}[ht]
\centering
\caption{Mapping between the key-mask pair and the corresponding decoded signals sent to the MvCAM cell.}
\label{tab:general_decoder}
\begin{tabular}{|c|c|ccccc|}
\hline
\multirow{2}{*}{Mask} & \multirow{2}{*}{Key} & \multicolumn{5}{c|}{Decoded signals}  \\
 & & $S_{n-1}$ & $S_{n-2}$ & $\cdots$ & $S_1$ & $S_0$ \\ \hline
0 & x& 0   & 0   & $\cdots$ & 0  & 0  \\
$n-1$   & 0& $n-1$  & $n-1$  & $\cdots$ & $n-1$ & 0  \\
$n-1$   & 1& $n-1$  & $n-1$  & $\cdots$ & 0  & $n-1$ \\
$\vdots$ & $\vdots$   & $\vdots$  & $\vdots$  & $\ddots$ & $\vdots$ & $\vdots$ \\
$n-1$   & $n-2$  & $n-1$  & 0   & $\cdots$ & $n-1$ & $n-1$ \\
$n-1$ & $n-1$ & 0 & $n-1$ & $\cdots$ & $n-1$ & $n-1$ \\ \hline
\end{tabular}
\vspace{-0.1in}
\end{table}


\subsection{MvCAM Array}
A MvCAM array consists of several MvCAM rows. A MvCAM row contains several ``$n$T$n$R'' cells along with a sensing circuit to distinguish between the full match and the mismatch states. A full match state is obtained when all cells in the row match the searched nits, while a mismatch is obtained when at least one cell in the row does not contain the searched nit. An essential requirement for the array is to concurrently compare the stored data in all rows with an inputted key and mask pair. For purposes of in-memory compute, we overwrite the activated columns of the matched rows with new data. The key determines the nit value to be searched for, while the mask determines the columns of interest to be separately activated during each of the compare and write operations. The nit key and its binary mask are inputted to a decoder that generates the corresponding signal vector $(S_{n-1},\, S_{n-2}\, ...\, S_1,\, S_0)$, as indicated in Table \ref{tab:general_decoder}. 
 
\subsubsection{Compare} 
\begin{figure}
 \centering
\includegraphics[width=1\linewidth,height=0.57\linewidth]
{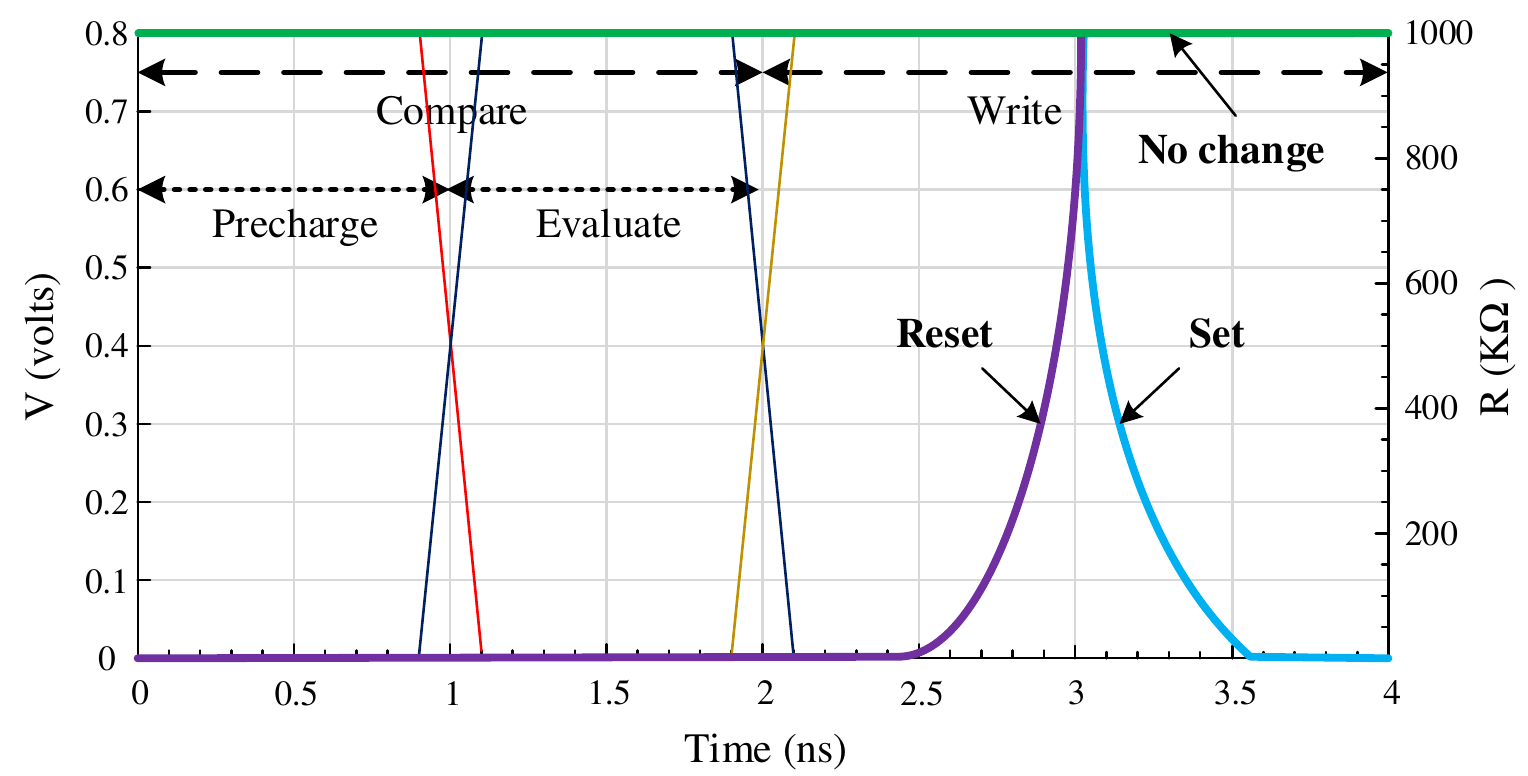}
 \caption{Sketch showing consecutive compare and write operations and the resulting memristor state changes for a ``3T3R'' CAM cell, assuming the compare operation results in a match.}
 \label{fig:timing_diagram}
 \vspace{-0.1in}
\end{figure}

The compare operation includes a precharge and an evaluate phase (see Fig. \ref{fig:timing_diagram}). During precharge, the capacitor is charged high, then a masked key is applied to the array in the evaluate phase. This leads the capacitor of each MvCAM row of cells to discharge through a resistor whose value is equal to the equivalent resistance of the corresponding row. In the case of a full match (fm), the capacitor retains most of its charges due to the presence of only high-resistance paths. In the case of one mismatching cell per row (1mm) or more (2mm, 3mm, etc.), the capacitor discharges quickly to ground through one, two or more low-resistance paths. 

\subsubsection{Write} 
After the compare operation, the sense amplifier connected to the output of the matching circuit senses the voltage across the capacitor and generates correspondences between a row match and logic `1', and a row mismatch and logic `0'. Hence, all matching rows are ``tagged'', meaning their Tag field is set to logic `1'. For example, in Fig. \ref{fig:mvap}, cells in the first row match the masked key and the row is tagged, whereas cells in the last row do not match the masked key. 
We note that the sensing amplifier is followed by a latch that holds the Tag bit throughout the write action. The write enable signal is asserted to overwrite the new masked columns of the tagged MvCAM rows with new data.
Each write action for an ``$n$T$n$R'' cell triggers one memristor set and one memristor reset, except for writing to (from) a ``don't care'' state which only requires one reset (set). This is attributed to the fact that each stored nit is associated with a distinct memristor set to $R_{LRS}$, except for the ``don't care'' value in which no memristor is set to $R_{LRS}$.

Finally, we note that in an optimized architecture, the precharge phase can be performed in parallel with the write operation to reduce the combined compare and write cycle time. As such, a transmission gate powered by the write enable signal is used to relay to the ML the proper programming voltage to program the memristors, while another pass gate powered by the inverse of the write enable isolates the precharge capacitor from the programming voltage, as illustrated in Fig. \ref{fig:mvap}. The ML programming voltage is set to high during reset, and to low during set. The ``$n$T$n$R'' virtual ground is set to zero during reset, and pulled high during set. 
 
\section{Illustrative TAP Architecture}
\label{sec:TAP}

In \cite{hayes2001third}, the author analyzed the number system to find the best radix in terms of the economical perspective (i.e., number of computations). The optimal radix is found to be the natural number $e=2.718$ \cite{radwan2015mathematical}. So, the ternary logic system is adopted as the best number system since the integer 3 is the nearest to $e$. Herein, we illustrate the MvAP architecture with a ternary AP (TAP) relying on a ``3T3R'' quaternary CAM (QCAM) cell. The ``3T3R'' cell is built using three transistors and three memristors and stores the ternary logic values `0', `1' and `2' in addition to a ``don't care'' value. 
Trits are stored in the form of one memristor set to $R_{LRS}$ and two memristors set to $R_{HRS}$. For example, the combination $(M_2,\, M_1,\, M_0) = (R_{HRS},\, R_{HRS},\, R_{LRS})$ indicates a logic `0' since $M_0$ is the memristor which is set to $R_{LRS}$ as shown in Table \ref{tab:tap_logic}. For the same stored value, when the decoded signal triplet sent is $(S_2,\, S_1,\, S_0) = (2,\, 2,\, 0)$, ML discharges very slowly since only high-resistance paths will connect ML to ground either through $R_{HRS}$ or $R_{off}$, thus resulting in a match. For all other decoded signal combinations, ML discharges quickly to ground through a low-resistance path, thus resulting in a mismatch.


\begin{table}[]
\centering
\caption{Different combinations of searched and stored data which can lead to either a match or a mismatch state. ``H'' and ``L'' donate $R_{HRS}$ and $R_{LRS}$, respectively.}
\label{tab:tap_logic}
\begin{tabular}{|c|c|c|ccc|c}
\cline{1-6}
\multicolumn{2}{|c|}{Searched Data} & \multicolumn{4}{c|}{Stored Data}  &  \\ \hline
Mask & Key & \begin{tabular}[c]{@{}c@{}}Logic\\ value\end{tabular} & $M_2$ & $M_1$ & $M_0$ & \multicolumn{1}{c|}{State} \\ \hline
0 & x & x & x & x & x & \multicolumn{1}{c|}{Match} \\
2 & 0 & 0 & H & H & L & \multicolumn{1}{c|}{Match} \\
2 & 1 & 0 & H & H & L & \multicolumn{1}{c|}{Mismatch} \\
2 & 2 & 0 & H & H & L & \multicolumn{1}{c|}{Mismatch} \\
2 & 0 & 1 & H & L & H & \multicolumn{1}{c|}{Mismatch} \\
2 & 1 & 1 & H & L & H & \multicolumn{1}{c|}{Match} \\
2 & 2 & 1 & H & L & H & \multicolumn{1}{c|}{Mismatch} \\
2 & 0 & 2 & L & H & H & \multicolumn{1}{c|}{Mismatch} \\
2 & 1 & 2 & L & H & H & \multicolumn{1}{c|}{Mismatch} \\
2 & 2 & 2 & L & H & H & \multicolumn{1}{c|}{Match} \\
2 & 0 & x & H & H & H & \multicolumn{1}{c|}{Match} \\
2 & 1 & x & H & H & H & \multicolumn{1}{c|}{Match} \\
2 & 2 & x & H & H & H & \multicolumn{1}{c|}{Match} \\ \hline
\end{tabular}
\end{table}

\begin{figure}
 \centering
\includegraphics[width=0.6\linewidth,height=0.6\linewidth]
{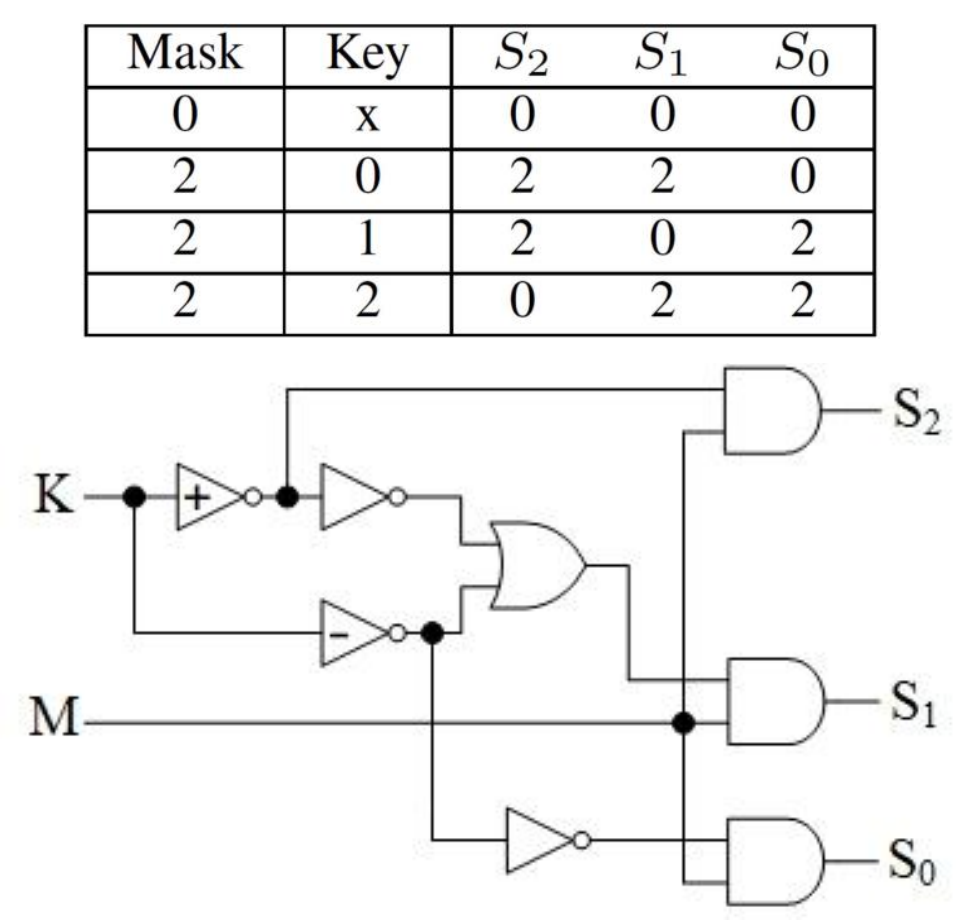}
 \caption{Ternary decoder; truth table (top) and circuit implementation (bottom). The inverters with `$+$' sign and `$-$' sign represent PTI and NTI, respectively. The rest of the gates are conventional binary gates }
 \label{fig:ternary_decoder}
 \vspace{-0.15in}
\end{figure}

\begin{table}[]
\centering
\vspace{-0.15in}
\caption{Truth table for the STI, PTI and NTI ternary inverters.}
\label{tab:inverter}
\begin{tabular}{|c|c|c|c|}
\hline
\textit{x} & \textit{STI(x)} & \textit{PTI(x)} & \textit{NTI(x)} \\ \hline
0& 2  & 2  & 2  \\ \hline
1& 1  & 2  & 0  \\ \hline
2& 0  & 0  & 0  \\ \hline
\end{tabular}
\end{table}

For the TAP, as is the case for the MvAP, the Key register contains the ternary values to be searched for inside the QCAM array, while the Mask register indicates which column or columns of the array are activated during comparison or writing. Upon compare, each key-mask pair generates a decoded signal triplet in which only one of the signals is set to zero, while the others are set to $V_{DD}$, i.e., logic value $n-1=2$ for the case of ternary logic. For example, to search for logic `0', the decoded signal triplet is $(S_2,\, S_1,\, S_0) = (2,\, 2,\, 0)$ as shown in Fig. \ref{fig:ternary_decoder}. Equations (\ref{eq:s2}), (\ref{eq:s1}) and (\ref{eq:s0}) represent the corresponding logic functions for the signal values obtained based on the truth table of Fig. \ref{fig:ternary_decoder}. The figure also presents the decoder circuit for the case of ternary logic comprising positive ternary inverters (PTIs), negative ternary inverters (NTIs) \cite{ternary_logic_gates}, binary AND, binary OR and binary inverter gates. The truth tables for the ternary inverters are depicted in Table \ref{tab:inverter} \cite{ternary_logic_gates}.  

 \begin{subequations}
 \begin{equation}
  S_2=Mask \cdot PTI(Key)
  \label{eq:s2}
 \end{equation}
 \begin{equation}
  S_1=Mask \cdot (NTI(Key) + \overline{PTI(Key)})
  \label{eq:s1}
 \end{equation}
 \begin{equation}
 S_0=Mask \cdot \overline{NTI(Key)}
\label{eq:s0}
 \end{equation}
 \end{subequations} 
For purposes of in-memory compute, matching rows are overwritten by new data. Each write operation of a ternary logic value includes one memristor set and one memristor reset, except for writing to (from) a ``don't care'' state which only requires one memristor reset (set).

To implement a specific ternary arithmetic function, we iterate through the LUT entries for 1-trit operation, and the process is repeated to perform multi-trit operations. For each 1-trit operation, the Key register is set to the corresponding LUT input values and applied concurrently to all rows of the array in the columns specified by the Mask register. These represent the operand columns. 
Each key-mask pair is fed to a decoder that generates the signal triplet $(S_2,\, S_1,\, S_0)$. After the compare operation, the rows of the QCAM array will generate either a match or a mismatch. Then, the write operation is performed on the matching rows of the array. New data consisting of the LUT output for the corresponding input replaces the stored value in the newly masked columns of the array. Table \ref{tab:set_reset} illustrates an example where trit $B$ initially takes on ternary value `1' which is equivalent to initial memristor states $(M_2,\, M_1,\, M_0) = (H,\, L,\, H)$. Assuming that the function's LUT dictates that $B$ should be overwritten by ternary value `0' post the compare operation, the final memristor states will be $(M_2,\, M_1,\, M_0) = (H,\, H,\, L)$. As such, for this example, $M_1$ should be reset and $M_0$ should be set, whereas $M_2$ remains the same as illustrated in Fig. \ref{fig:timing_diagram}.

\begin{table}[t]
\centering
\caption{A write example for the ternary ``3T3R'' cell. `x', `R' and `S' mean no change, reset and set, respectively. }
\label{tab:set_reset}
\begin{tabular}{l|c|c|c|}
\cline{2-4}
  & $A$ & $B$ & $C_{in}$ \\ \hline
\multicolumn{1}{|l|}{Current state} & 0 & 1 & 2  \\ \hline
\multicolumn{1}{|l|}{Current stored ($M_2$, $M_1$, $M_0$)} & (H, H, L) & (H, L, H) & (L, H, H) \\ \hline
\multicolumn{1}{|l|}{Next state} & 0 & 0 & 1  \\ \hline
\multicolumn{1}{|l|}{Next stored ($M_2$, $M_1$, $M_0$)} & (H, H, L) & (H, H, L) & (H, L, H) \\ \hline
\multicolumn{1}{|l|}{Action}  & (x, x, x) & (x, R, S) & (R, S, x) \\ \hline
\end{tabular}
\vspace{-0.15in}
\end{table}

\section{AP Operation}
\label{sec:AP_Operation}
A general-purpose AP enables the implementation of arithmetic functions such as addition, subtraction, multiplication and division as well as logical operations by relying on the truth tables of the desired function applied in a specific order. We refer to this as the look-up table based approach. The process is performed digit-wise and is repeated for multi-digit operations. The rows of the MvCAM array store the input vectors. For in-place operation, the output is written back to some or all of the input locations. All rows of the data array are processed in parallel. Each digit-wise operation is comprised of consecutive compare and write steps.
\begin{enumerate}
 \item \textbf{Compare:} For every pass of the LUT, a masked key takes on the input vector values of this pass (see Table \ref{tab:binary_LUT} for the example of binary AP addition \cite{binary_AP}). The masked key is applied to all rows of the array and compared against the stored input data.
 \item \textbf{Write:} A match for a row sets its Tag bit to a `1', while a mismatch for a row sets its Tag bit to a `0'. Tagged rows are overwritten by the corresponding output from the LUT consisting of a new masked key. For the example of the binary AP, the sum bit $S$ and the carry-out bit $C_{out}$ are written back to the input locations $B$ and $C_{in}$ respectively, keeping $A$ untouched. The in-place write-back of the output dictates the order of the passes in the LUT. This is required to avoid mistakenly revisiting in future passes rows that have already been overwritten, as will be discussed later. 
\end{enumerate}

\begin{table}[b]
\vspace{-0.15in}
\centering
\caption{Look-up table of the binary AP adder.}
\label{tab:binary_LUT}
\begin{tabular}{|ccc|ccc|c|}
\hline
\multicolumn{3}{|c|}{Input} & \multicolumn{3}{c|}{Output} & \multirow{2}{*}{\begin{tabular}[c]{@{}c@{}}Pass\\ order\end{tabular}} \\
A & B & C & A & B & C & \\ \hline
0 & 0 & 0 & 0 & 0 & 0 & No action \\
0 & 0 & 1 & 0 & 1 & 0 & 3   \\
0 & 1 & 0 & 0 & 1 & 0 & No action \\
0 & 1 & 1 & 0 & 0 & 1 & 4   \\
1 & 0 & 0 & 1 & 1 & 0 & 2   \\
1 & 0 & 1 & 1 & 0 & 1 & No action \\
1 & 1 & 0 & 1 & 0 & 1 & 1   \\
1 & 1 & 1 & 1 & 1 & 1 & No action \\ \hline
\end{tabular}
\end{table}

\subsection{Proposed State Diagram for LUT Generation}
The proper pass order for a given arithmetic function can be ensured as follows. Consider that $x$ appears in the truth table of the function as both an input in one entry and an output in some other entry, then the order of processing $x$ as an input must satisfy one of the two properties below:
\begin{enumerate}
\item The pass in which $x$ appears as an input must be tested before the pass in which $x$ appears as an output. 
\item $x$ as an input results in `No action', i.e., the output to be overwritten is identical to the stored input. Such an input has no pass number because no action is needed, hence it will never be tested after the pass in which $x$ appears as an output. This implies that the order of the pass in which $x$ appears as an output is independent of the pass in which $x$ appears as an input.
\end{enumerate}
These properties ensure that the resulting passes are visited correctly. 

\begin{figure}
 \centering
\includegraphics[width=0.655\linewidth,height=0.65\linewidth]
{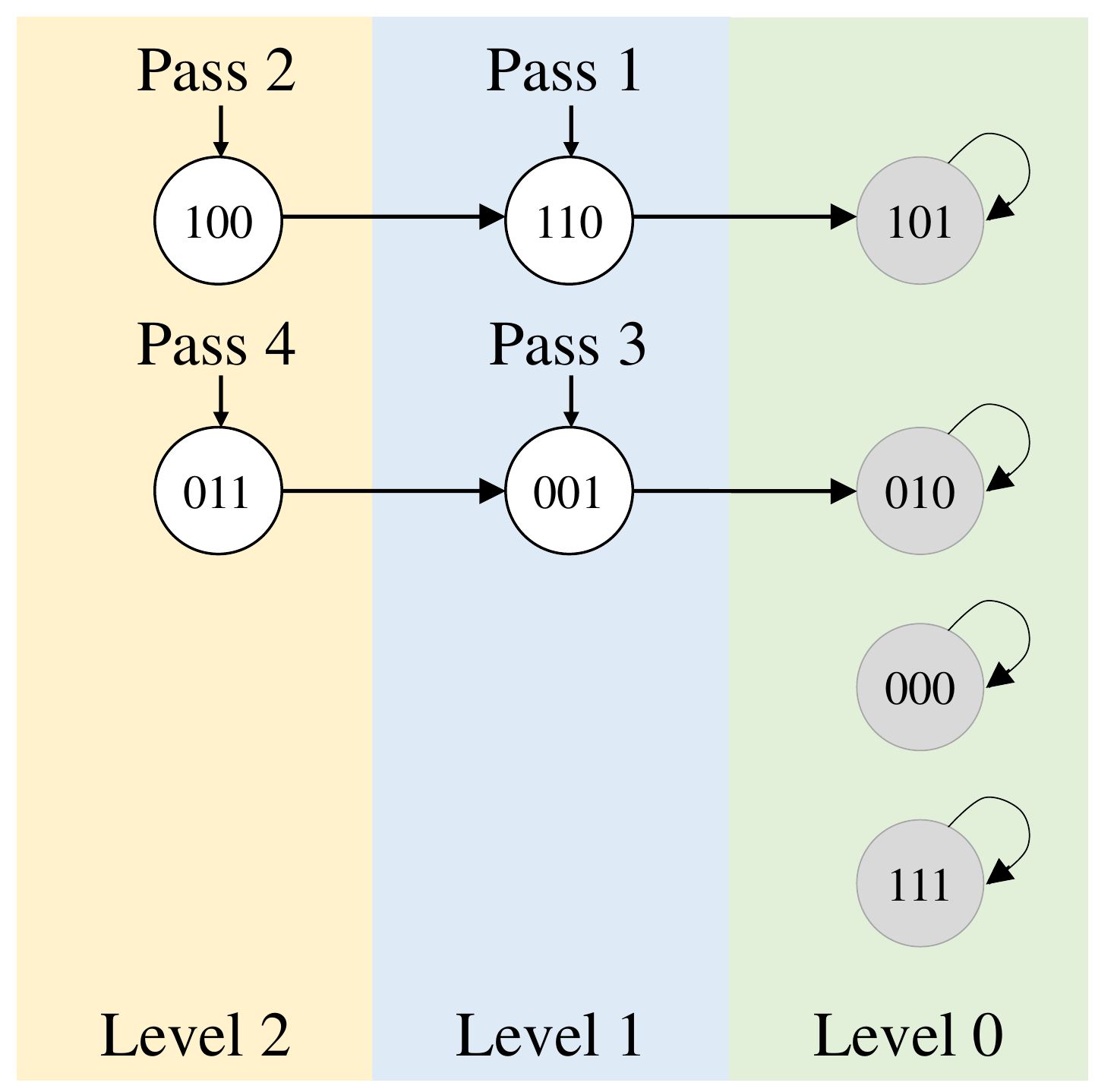}
 \caption{State diagram of the AP binary adder. The states store the values for the triplets ($A$, $B$, $C_{in}$). The arrow represents the addition operation. The pass order of Table \ref{tab:binary_LUT} are labeled.}
 \label{fig:binary_state_diagram}
 \vspace{-0.15in}
\end{figure}

Herein, we propose a directed state diagram representation of the truth table of the arithmetic or logic function to be implemented using AP. Our objective is to rely on this state diagram representation to identify the proper processing order of the function's truth table and, accordingly, generate its LUT. 
The elements of the state diagram can be best described as follows. 
\begin{itemize}
 \item Directed edge: application of the arithmetic function under consideration. 
 \item State: stored input to be operated upon.
 \item Next state: corresponding output as per the LUT.
 \item $noAction$ state: state that remains the same upon in-place operation, that is, the LUT input is identical to its corresponding LUT output.
\end{itemize}
Without loss of generality, Fig. \ref{fig:binary_state_diagram} first illustrates the state diagram of the binary adder's truth table. In this example, the edge corresponds to the binary add operation, the state corresponds to the 1-bit input triplet $(A_i,\, B_i,\, C_{in})$ and the next state corresponds to the output triplet $(A_i,\, S_i,\, C_{out})$. Finally, the $noAction$ state is the one pointing to itself, indicating that it remains the same upon in-place addition, that is, $(A_i,\, S_i,\, C_{out}) = (A_i,\, B_i,\, C_{in})$. 
The binary adder state diagram in Fig. \ref{fig:binary_state_diagram} is also labeled with the LUT pass order \cite{binary_AP}. An analysis of the pass order shows that they were ordered to avoid erroneously performing multiple consecutive additions on the same entry. This translates in the state diagram to the first pass writing `110' by `101', and the second pass writing `100' by `110'. No other pass can overwrite these outputs once their respective inputs are visited. On the other hand, if passes 1 and 2 are exchanged, `100' results in `110' after the first pass, which will be overwritten by `101' after the second pass as indicated by the directional flow of the state diagram. Such a domino effect is not desired. Therefore, it is evident that to construct the LUT for a generic arithmetic function, the order of the passes must be determined through a structured traversal of the directed state diagram.

\subsection{Automated LUT Generation}
\begin{figure}[t]
 \centering
\includegraphics[width=1\linewidth, height=0.85\linewidth]
{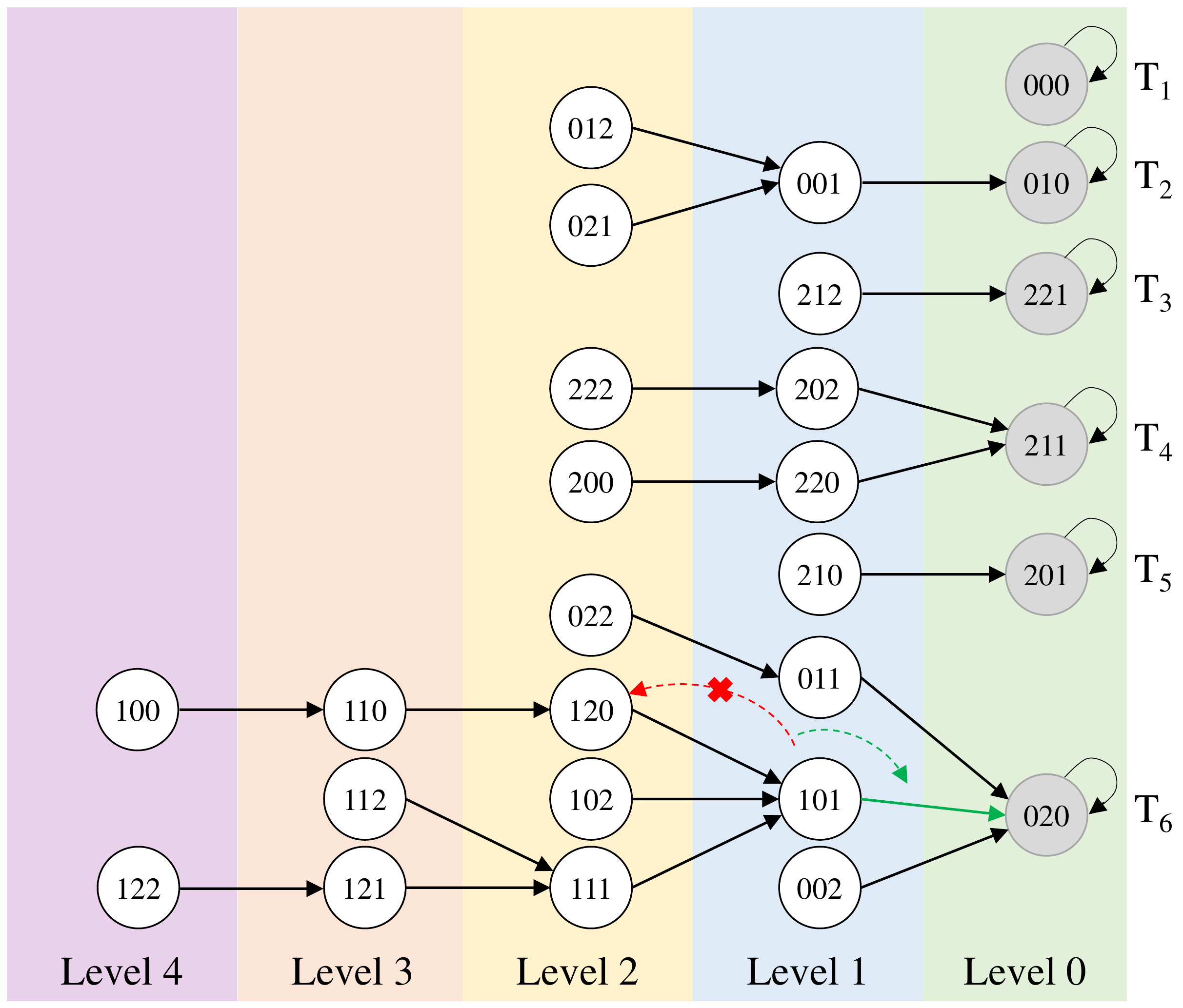}
 \caption{State diagram of AP ternary full adder. For each subtree, the passes are determined based on a depth-first search to enforce the priorities.}
 \label{fig:TFA_state_diagram}
 \vspace{-0.15in}
\end{figure}

Herein, we build upon our state diagram interpretation of the truth table to guide the automatic development of a general-purpose LUT. As we note from Fig. \ref{fig:binary_state_diagram}, the state diagram comprises of a collection of trees whose roots are $noAction$ states. We note that the input-output pairs are connected through backward edges propagating to the roots. Our objective is to identify the proper order of passes for in-place operation so that no pass overwrites the outcome of earlier ones. This can be guaranteed if and only if the following holds for the state diagram.
\begin{enumerate}
 \item The state diagram is a uni-directional graph with no cycles, i.e., no forward edges.
 \item If the state diagram has cycles, then we should be able to break these cycles by redirecting forward edges backwards. If in the original state diagram input vector $x = (x_1,\, x_2)$ has its output $y = (x_1,\, y_2)$ creating a forward edge, we search for an alternate output $y' = (y_1,\, y_2)$ that forms a backward edge and breaks the cycle. $y'$ is a valid output so long $y_2$ remains unchanged since $y_2$ represents the output digit to be overwritten as per the LUT, while $y_1$ is a dummy extra written digit. Therefore, we need to invoke a larger vector write post the compare operation. This will be illustrated in the following example. 
\end{enumerate}
For example, we implement the state diagram for the LUT-based ternary full adder (TFA) in the context of TAP. Fig. \ref{fig:TFA_state_diagram} presents the state diagram of the TFA's truth table. We perform in-place ternary addition with inputs $(A,\, B,\, C_{in})$. The outputs $(S,\, C_{out})$ overwrite $(B,\, C_{in})$, while $A$ is kept untouched. In the state diagram, if the input triplet $(A,\, B,\, C_{in})$ whose output $(A,\, S,\, C_{out})$ represents a forward edge forming a cycle, we search for an alternate output $(y_1,\, S,\, C_{out})$ that forms a backward edge and breaks the cycle. We then invoke a 3-trit write post the compare operation instead of the standard 2-trit write. Specifically, as illustrated by the dashed red edge in Fig. \ref{fig:TFA_state_diagram}, a direct implementation of the in-place ternary addition state diagram results in one cycle: state `101' leads to `120' and state `120' leads back to `101'. To resolve this problem, we overwrite the $A$ trit value to a `0’ for the input `101'. Hence, input `101' now results in `020' as an output (see green edge in Fig. \ref{fig:TFA_state_diagram}) and input `120' results in `101' as an output. This incurs a minor cost consisting of an extra trit to be written for one of the passes. However, it eliminates the cycle and enables a smooth implementation of the LUT-based approach for the TFA. 

With a cycle-free state diagram, i.e., a backward propagation based input-output relation (left to right), we devise that the passes should progress to visit the trees of the state machine from right to left in a depth-first search (DFS) approach, starting from the root of each tree. Since the roots are $noAction$ nodes, we do not assign pass numbers to them and, hence, do not include them in the ordering of the passes. Algorithm \ref{alg:passes_order_nb} presented herein details the traversal scheme of the state diagram which ultimately determines the proper order of the passes for in-place operation. Table \ref{tab:ternary_nb_LUT} presents the resulting LUT for the TFA after applying Algorithm \ref{alg:passes_order_nb}.

\begin{algorithm}[h!]
\caption{Ordering of the passes for the LUT-based ternary full adder following the non-blocked approach.
\label{alg:passes_order_nb}}
\begin{algorithmic}[1]
 \STATE Global pass number $p=0$
 \STATE Global LUT length $L=length(LUT)$
 \FORALL{$T_i$}
  \item \textsc{BuildLUT}($T_i.root$)
 \ENDFOR
 \RETURN
\end{algorithmic}

\begin{algorithmic}[1]
 \item[]
 \item[\textbf{procedure}{ \textsc{BuildLUT}(state $j$)}]
  \IF{$j.noAction == 0$}
    \STATE $p++$
    \STATE $j.passNum=p$
  \ENDIF
  \FORALL{$v \in j.child$}
   \STATE \textsc{BuildLUT}($v$)
  \ENDFOR
  \RETURN
 \end{algorithmic}
\end{algorithm}

\begin{table}[]
\centering
\caption{Look-up table of the LUT-based TFA.}
\label{tab:ternary_nb_LUT}
\begin{tabular}{|ccc|ccc|c|}
\hline
\multicolumn{3}{|c|}{Input} & \multicolumn{3}{c|}{Output} & \multirow{2}{*}{\begin{tabular}[c]{@{}c@{}}Pass\\ order\end{tabular}} \\
A & B & C & A & B & C & \\ \hline
0 & 0 & 0 & 0 & 0 & 0 & No action \\
0 & 0 & 1 & 0 & 1 & 0 & 1   \\
0 & 0 & 2 & 0 & 2 & 0 & 21  \\
0 & 1 & 0 & 0 & 1 & 0 & No action \\
0 & 1 & 1 & 0 & 2 & 0 & 10  \\
0 & 1 & 2 & 0 & 0 & 1 & 2   \\
0 & 2 & 0 & 0 & 2 & 0 & No action \\
0 & 2 & 1 & 0 & 0 & 1 & 3   \\
0 & 2 & 2 & 0 & 1 & 1 & 11  \\
1 & 0 & 0 & 1 & 1 & 0 & 15  \\
1 & 0 & 1 & 0 & 2 & 0 & 12  \\
1 & 0 & 2 & 1 & 0 & 1 & 16  \\
1 & 1 & 0 & 1 & 2 & 0 & 14  \\
1 & 1 & 1 & 1 & 0 & 1 & 17  \\
1 & 1 & 2 & 1 & 1 & 1 & 18  \\
1 & 2 & 0 & 1 & 0 & 1 & 13  \\
1 & 2 & 1 & 1 & 1 & 1 & 19  \\
1 & 2 & 2 & 1 & 2 & 1 & 20  \\
2 & 0 & 0 & 2 & 2 & 0 & 8   \\
2 & 0 & 1 & 2 & 0 & 1 & No action \\
2 & 0 & 2 & 2 & 1 & 1 & 5   \\
2 & 1 & 0 & 2 & 0 & 1 & 9   \\
2 & 1 & 1 & 2 & 1 & 1 & No action \\
2 & 1 & 2 & 2 & 2 & 1 & 4   \\
2 & 2 & 0 & 2 & 1 & 1 & 7   \\
2 & 2 & 1 & 2 & 2 & 1 & No action \\
2 & 2 & 2 & 2 & 0 & 2 & 6   \\ \hline
\end{tabular}
\vspace{-0.1in}
\end{table}

\section{Optimized AP Operation}
\label{sec:Optimized_AP}

In the first approach, hereafter referred to as the \emph{non-blocked} approach, similar to traditional AP operation, each pass comprises of a compare cycle followed by a write cycle. Often, for a given function, different input vectors result in similar output vectors. For example, for the TFA, different input triplets $(A,\, B,\, C_{in})$ often result in similar output pairs $(S,\, C_{out})$. Since write cycles are much more expensive than compare cycles, we can leverage this fact to devise a second approach that capitalizes on these common output vectors. 
As such, we propose a more optimized approach that targets the dual objective of determining the proper pass order and grouping different input vectors that share the same output vector. The proposed approach starts by processing compare cycles for the input vectors that have the same output vector, then the write action occurs once all input vectors of a group (or block) are visited, thereby improving the efficiency of the LUT-based approach. Hereafter, we refer to this as the \emph{blocked} approach. 

In order to traverse the state diagram using the blocked approach in such a way to determine the correct ordering of the passes, we adopt a breadth-first search (BFS)-like traversal of the nodes. Every time a new block is determined, the state diagram is dynamically updated to eliminate the most recently chosen block nodes. Post each update, we group nodes to the same block in terms of (a) children of the same node, and (b) other nodes at the same level sharing the same write action. Note that the set of nodes (b) may not have necessarily existed in the initial state diagram at the same level as set of nodes (a); however, dynamic updates of the state diagram as we construct the LUT will enforce such conditions as we explain next. 

To implement the algorithm for the automatic generation of the LUT using the blocked approach, we consider that each node in the state diagram represents an input or output vector depending on whether the node is subject to or the result of an add operation, respectively. Each node is also associated with a set of attributes detailed in Table \ref{tab:LUT_definitions}. One of the attributes is $writeDim$ representing the dimension of the output vector to be written when the node is regarded as an input state. Another attribute is the $outVal$ array whose entries represent the `$n$-ary'-to-decimal value of the node's write action when it is regarded as an output state. The need for the $outVal$ array of entries is explained as follows using the TFA as an example. For the TFA, when the node is the result of a 2-trit write-back, $outVal(writeDim=2)$ stores the ternary-to-decimal conversion of the written $BC$ value. In the event of breaking cycles, we may add an extra dummy dimension and invoke a 3-trit write-back. The node that is regarded as the input of the operation will have $writeDim=3$ and the node that is regarded as the output will have $outVal(writeDim=3)$ store the corresponding equivalent decimal written $ABC$ value. Note that these values will be adjusted as explained later to avoid overlap between the different decimal value conversions. For example, $ABC = 000$ will be mapped to a different number than $BC = 00$ as indicated in line 5 of Algorithm \ref{alg:filling_grplvl}. This will help in properly differentiating grouping of nodes that have the same parent but different write action dimensions. 
Thus, nodes at the same level having the same $writeDim$ and parent $outVal$ share the same write action. As such, for a specific node, we rely on its parent $outVal(writeDim)$ value for the grouping and use it as a key component of the dynamic $grpLvl$ table that we rely on to guide the BFS-like traversal algorithm. The table stores the number of nodes belonging to the same group, i.e., having a similar write action, in each level of the tree. The algorithm performs pass ordering and grouping by updating the $grpLvl$ table after each determined block to reflect updates to the state diagram. The algorithm proceeds as follows. 

\begin{table}[!t]
\centering
\caption{State attributes definitions for the blocked and non-blocked algorithms.}
\label{tab:LUT_definitions}
\begin{tabular}{|l|l|}
\hline
\multicolumn{1}{|c|}{Attribute} & \multicolumn{1}{c|}{Definition}       \\ \hline
$noAction$ & \begin{tabular}[c]{@{}l@{}}Determines the type of the state:\\ 1 for a No Action state\\ 0 for an Action state\end{tabular}    \\ \hline
$grpNum$      & Specific write group of the state      \\ \hline
$level$ & State level as indicated in Fig. \ref{fig:TFA_state_diagram}               \\ \hline
$outVal$      & `$n$-ary'-to-decimal conversion of the state vector  \\ \hline
$writeDim$      & Write-back dimension for the output vector of the state \\ \hline
$parent$ & \begin{tabular}[c]{@{}l@{}}Pointer to the parent of the state which is accessible from\\ the state through a backward edge\end{tabular} \\ \hline
$child$ & \begin{tabular}[c]{@{}l@{}}Pointer to the child of the state which is accessible from\\ the state through a forward edge\end{tabular} \\ \hline
$passNum$      & Pass order assigned to the state in the LUT    \\ \hline
\end{tabular}
\vspace{-0.15in}
\end{table}

\subsubsection{$grpLvl$ initialization} 
To populate the initial $grpLvl$ table, we apply Algorithm \ref{alg:filling_grplvl}.
Table \ref{tab:grp_lvl} represents the initial $grpLvl$ table for the TFA state diagram corresponding to Fig. \ref{fig:TFA_state_diagram}. 
For each state which is an $Action$ state, we find $l$, the level of the node as indicated by Fig. \ref{fig:TFA_state_diagram}, and $g$, the $outVal(writeDim)$ value of the node's parent. For example, the group number $g$ for node `101' in the TFA state diagram of Fig. \ref{fig:TFA_state_diagram} is $outVal(3)$ of its parent node `020'. This corresponds to the adjusted value $6+\sum_{i=0}^{2}3^i=19$, where `6' is the ternary-to-decimal conversion of the vector `020'. Whereas the group number $g$ for node `011' is $outVal(2)$ of its parent node `020', corresponding to the adjusted value $6+\sum_{i=0}^{1}3^i=10$, where `6' is the ternary-to-decimal conversion of the vector `20'. Accordingly, for each $Action$ node in level $l$, an initial group number $grpNum=g$ is assigned to the node, and the entry corresponding to group $g$ and level $l$ in the $grpLvl$ table is incremented. Entries in the $grpLvl$ table thus reflect the number of nodes that share the same level and write action. For example, 5 nodes in Level 2 share the same write action $BC=01_3=1_{10}$, having an adjusted value of $1+\sum_{i=0}^{1}3^i=5$. Thus, $grpLvl[l=2][g=5]=5$ as shown in Table \ref{tab:grp_lvl}. 

\begin{algorithm}[h!]
\caption{Initializing the $grpLvl$ table.}
\label{alg:filling_grplvl}
\begin{algorithmic}[1]
\item [Global pass number $p=0$]
\item [\# Preparing Action states group table $grpLvl$]
\STATE $S$ is the set of all states $\forall T_i$
 \FORALL{states $j \in S$}
  \IF{$j.noAction==0$}
   \STATE Level $l=j.level$
   \STATE Group number $g=j.parent.outVal(writeDim)+\sum_{i=0}^{writeDim-1}(n^{i})$ 
   \STATE $j.grpNum=g$
   \STATE $grpLvl[l][g]++$
  \ENDIF
 \ENDFOR

 \STATE $G=max({g})$
 \STATE $L=max({l})$
 \item[\# Use $grpLvl$ table to build LUT for Action states] 
 \STATE
 \textsc{BuildLUTBlocked($S$, $grpLvl$, $G$, $L$)}
\RETURN
\end{algorithmic}
\end{algorithm}

\begin{table*}[]
\centering
\caption{$grpLvl$ table initial values corresponding to Fig. \ref{fig:TFA_state_diagram}. It indicates that Group 19 should be processed first since it is the only group that has no entries beyond Level 1. Note that for the TFA example, $writeDim=1$ does not exist, and thus by default, no nodes can have $grpNum=\{1,\, 2,\, 3\}$.}
\label{tab:grp_lvl}
\begin{tabular}{cc|c|c|c|c|c|c|c|c|c|c|c|c|c|c|c|c|c|c|c|}
\cline{3-21}
 &
 &
 \multicolumn{19}{c|}{\cellcolor[HTML]{EFEFEF}$grpNum$} \\ \cline{3-21} 
 &
 &
 \multicolumn{3}{c|}{\begin{tabular}[c]{@{}c@{}}$parent.$\\ $outVal(1)$\\ $+\sum_{i=0}^{0}3^i$\end{tabular}} &
 \multicolumn{9}{c|}{$parent.outVal(2)+\sum_{i=0}^{1}3^i$} &
 \multicolumn{7}{c|}{$parent.outVal(3)+\sum_{i=0}^{2}3^i$} \\ \cline{3-21} 
 &
 &
 \textbf{1} &
 \textbf{2} &
 \textbf{3} &
 \textbf{4} &
 \textbf{5} &
 \textbf{6} &
 \textbf{7} &
 \textbf{8} &
 \textbf{9} &
 \textbf{10} &
 \textbf{11} &
 \textbf{12} &
 \textbf{13} &
 \textbf{14} &
 \textbf{15} &
 \textbf{16} &
 \textbf{17} &
 \textbf{18} &
 \textbf{19} \\ \hline
\multicolumn{1}{|c|}{\cellcolor[HTML]{EFEFEF}} &
 \textbf{1} &
 0 &
 0 &
 0 &
 0 &
 {\color[HTML]{DD0808} 1} &
 0 &
 {\color[HTML]{DD0808} 1} &
 {\color[HTML]{DD0808} 2} &
 0 &
 {\color[HTML]{DD0808} 2} &
 {\color[HTML]{DD0808} 1} &
 0 &
 0 &
 0 &
 0 &
 0 &
 0 &
 0 &
 {\color[HTML]{DD0808} 1} \\ \cline{2-21} 
\multicolumn{1}{|c|}{\cellcolor[HTML]{EFEFEF}} &
 \textbf{2} &
 0 &
 0 &
 0 &
 0 &
 {\color[HTML]{DD0808} 5} &
 {\color[HTML]{DD0808} 1} &
 0 &
 {\color[HTML]{DD0808} 1} &
 0 &
 {\color[HTML]{DD0808} 1} &
 0 &
 0 &
 0 &
 0 &
 0 &
 0 &
 0 &
 0 &
 0 \\ \cline{2-21} 
\multicolumn{1}{|c|}{\cellcolor[HTML]{EFEFEF}} &
 \textbf{3} &
 0 &
 0 &
 0 &
 0 &
 0 &
 0 &
 0 &
 {\color[HTML]{DD0808} 2} &
 0 &
 {\color[HTML]{DD0808} 1} &
 0 &
 0 &
 0 &
 0 &
 0 &
 0 &
 0 &
 0 &
 0 \\ \cline{2-21} 
\multicolumn{1}{|c|}{\multirow{-4}{*}{\cellcolor[HTML]{EFEFEF}$level$}} &
 \textbf{4} &
 0 &
 0 &
 0 &
 0 &
 0 &
 0 &
 {\color[HTML]{DD0808} 1} &
 0 &
 0 &
 0 &
 {\color[HTML]{DD0808} 1} &
 0 &
 0 &
 0 &
 0 &
 0 &
 0 &
 0 &
 0 \\ \hline
\end{tabular}
\vspace{-0.15in}
\end{table*}

\subsubsection{Selecting the next block/group}
At each iteration, our objective is to find the next target block/group $g_{tgt}$ for an adequate ordering of the passes in the LUT. We keep in mind the following. Nodes that reside in Level 1 must be processed first from a pass ordering perspective (these qualify as nodes whose parents have already been processed or whose parents are $noAction$ states). For purposes of grouping, we must therefore look for groups that are fully or maximally residing in Level 1. In fact, we consider the following two cases as indicated in Algorithm \ref{alg:finding_gtgt}. 
\begin{itemize}
 \item In an ideal scenario, there exists a group that has nodes belonging to the top level (Level 1) and no nodes in lower levels. This group would qualify as the next target group $g_{tgt}$. 
 \item Another possible scenario is that no group has all its nodes in the top level. In this case, we choose the group that has a maximum number of nodes in the top level as $g_{tgt}$. However, we need to break this group since we can only process states belonging to the top level. We split $g_{tgt}$ by creating a new group for the remaining states present in lower levels. In this way, the total number of groups $G$ is incremented, and $g_{tgt}$ will only contain nodes that are in the top level. 
\end{itemize}
To illustrate, initial $grpLvl$ values shown in Table \ref{tab:grp_lvl} indicate that Group 19, should be processed first. It is the only group that has entries in the top level and no entries in lower levels. Supplementary Tables 1, 2 and 3 present $grpLvl$ tables for the following three iterations, and at each iteration, we identify all possible new $g_{tgt}$ groups. We continue until all the entries in the $grpLvl$ table at the top level become zero. 

\begin{algorithm}
\caption{Finding the next block/group $g_{tgt}$.}
\label{alg:finding_gtgt} 
\begin{algorithmic}[1]
 \item[\textbf{procedure}{ \textsc{BuildLUTBlocked}($S$, $grpLvl$, $G$, $L$)}]
 \STATE $topLevel=1$
 \WHILE{$grpLvl[topLevel][.] \neq zeros(1,G)$}
  \STATE $found=-1$
  \FOR {$g=0 \to G$} 
   \STATE $cond_1=(grpLvl[topLevel][g]>0)$
   \STATE $cond_2=(\sum_{l=2}^{L}{grpLvl[l][g]} == 0)$
   \IF{$cond_1$ $\AND$ $cond_2$} 
    \STATE $g_{tgt}=g$
    \STATE \textsc{updateLUT($g_{tgt}$)}
    \STATE $found=1$
   \ENDIF
  \ENDFOR
  \IF{$found==-1$}
   \STATE $[g_{tgt}, max_{grpLvl}]=max(grpLvl[topLevel][.])$
   \item[\# Create new group for remaining states of $g_{tgt}$ in]
   \item[\# lower levels]
    \STATE $G++$
    \FOR{$l=2 \to L$}
    \STATE $grpLvl[l][G]=grpLvl[l][g_{tgt}]$
    \STATE $grpLvl[l][g_{tgt}]=0$
   \ENDFOR 
   \FORALL{states $j$}
   \IF{($j.grpNum==g_{tgt}$) $\AND$ ($j.level>1$)}
   \STATE $j.grpNum=G$   
   \ENDIF
   \ENDFOR
   \STATE \textsc{updateLUT}($g_{tgt}$)
  \ENDIF
 \ENDWHILE
\RETURN
\end{algorithmic}
\end{algorithm}

\subsubsection{Updating $grpLvl$ and assigning pass numbers}
We rely on Algorithm \ref{alg:passes_order_blocked} to order the passes and build the LUT. In each iteration, once the next $g_{tgt}$ is identified, we extract the nodes with $grpNum=g_{tgt}$ from the state diagram and assign them as the next block to be processed in the LUT. We label them accordingly with their corresponding pass number. Note that within the $g_{tgt}$ group, passes can be numbered arbitrarily. For example, in Group 2 shown in Table \ref{tab:ternary_blocked_LUT}, triplet `102' order can be interchanged with any other triplet order from the same group, say `120'. The children of the selected group nodes are then elevated to Level 1 and their subtree nodes are elevated by one level as well. $grpLvl$ table and the state diagram are updated to reflect the changes accordingly. Finally, we set the top level entry in the $grpLvl$ table corresponding to $g_{tgt}$ to zero. Hence, as the state diagram is traversed and pass numbers are allocated to the nodes, entries in the $grpLvl$ table will be updated, mimicking updates in the state diagram. This is exemplified by Supplementary Fig. 1, 2 and 3 and their corresponding $grpLvl$ Tables 1, 2 and 3. The resulting LUT for the TFA following the blocked approach is shown in Table \ref{tab:ternary_blocked_LUT}.


\begin{algorithm}
\caption{Updating the $grpLvl$ table and ordering the passes of the LUT.}
\label{alg:passes_order_blocked}
\begin{algorithmic}[1]
  \item[\textbf{procedure}{ \textsc{updateLUT}($g_{tgt}$)}]
  \STATE $topLevel=1$
 \item[\# Generate pass number for states in the target group]
 \FORALL{states $j$}
  \IF{$j.grpNum==g_{tgt}$}
   \STATE $p++$
   \STATE $j.passNum=p$
   \FORALL{$v \in$ tree whose root is $j$}
    \STATE $grpLvl[v.level-1][v.grpNum]++$
    \STATE $grpLvl[v.level][v.grpNum]--$
    \STATE $v.level--$
   \ENDFOR
  \ENDIF
 \ENDFOR
 \STATE $grpLvl[topLevel][g_{tgt}]=0$
\RETURN
\end{algorithmic}
\end{algorithm}

\begin{table}[]
\centering
\caption{Look-up table of the LUT-based TFA following the blocked approach.}
\label{tab:ternary_blocked_LUT}
\begin{tabular}{|ccc|ccc|c|c|c|}
\hline
\multicolumn{3}{|c|}{Input} &
 \multicolumn{3}{c|}{Output} &
 \multirow{2}{*}{\begin{tabular}[c]{@{}c@{}}Pass\\ order\end{tabular}} &
 \multirow{2}{*}{\begin{tabular}[c]{@{}c@{}}Group\\ number\end{tabular}} &
 \multirow{2}{*}{\begin{tabular}[c]{@{}c@{}}Write\\ action\end{tabular}} \\
A & B & C & A & B & C &   &      &      \\ \hline
1 & 0 & 1 & 0 & 2 & 0 & 1   & 1      & W020     \\ \hline
1 & 0 & 2 & 1 & 0 & 1 & 2   & \multirow{4}{*}{2} & \multirow{4}{*}{W01} \\
1 & 1 & 1 & 1 & 0 & 1 & 3   &      &      \\
1 & 2 & 0 & 1 & 0 & 1 & 4   &      &      \\
2 & 1 & 0 & 2 & 0 & 1 & 5   &      &      \\ \hline
1 & 1 & 2 & 1 & 1 & 1 & 6   & \multirow{4}{*}{3} & \multirow{4}{*}{W11} \\
1 & 2 & 1 & 1 & 1 & 1 & 7   &      &      \\
2 & 0 & 2 & 2 & 1 & 1 & 8   &      &      \\
2 & 2 & 0 & 2 & 1 & 1 & 9   &      &      \\ \hline
0 & 0 & 2 & 0 & 2 & 0 & 10  & \multirow{4}{*}{4} & \multirow{4}{*}{W20} \\
0 & 1 & 1 & 0 & 2 & 0 & 11  &      &      \\
1 & 1 & 0 & 1 & 2 & 0 & 12  &      &      \\
2 & 0 & 0 & 2 & 2 & 0 & 13  &      &      \\ \hline
1 & 2 & 2 & 1 & 2 & 1 & 14  & \multirow{2}{*}{5} & \multirow{2}{*}{W21} \\
2 & 1 & 2 & 2 & 2 & 1 & 15  &      &      \\ \hline
0 & 0 & 1 & 0 & 1 & 0 & 16  & \multirow{2}{*}{6} & \multirow{2}{*}{W10} \\
1 & 0 & 0 & 1 & 1 & 0 & 17  &      &      \\ \hline
2 & 2 & 2 & 2 & 0 & 2 & 18  & 7      & W02     \\ \hline
0 & 1 & 2 & 0 & 0 & 1 & 19  & \multirow{2}{*}{8} & \multirow{2}{*}{W01} \\
0 & 2 & 1 & 0 & 0 & 1 & 20  &      &      \\ \hline
0 & 2 & 2 & 0 & 1 & 1 & 21  & 9      & W11     \\ \hline
0 &
 0 &
 0 &
 0 &
 0 &
 0 &
 No action &
 \multicolumn{1}{l|}{\multirow{6}{*}{}} &
 \multirow{6}{*}{} \\
0 & 1 & 0 & 0 & 1 & 0 & No action & \multicolumn{1}{l|}{} &      \\
0 & 2 & 0 & 0 & 2 & 0 & No action & \multicolumn{1}{l|}{} &      \\
2 & 0 & 1 & 2 & 0 & 1 & No action & \multicolumn{1}{l|}{} &      \\
2 & 1 & 1 & 2 & 1 & 1 & No action & \multicolumn{1}{l|}{} &      \\
2 & 2 & 1 & 2 & 2 & 1 & No action & \multicolumn{1}{l|}{} &      \\ \hline
\end{tabular}
\vspace{-0.15in}
\end{table}


\subsubsection*{\textbf{Circuits to Enable the Blocked Approach}} We note that there is a minimal cost overhead for blocking to delay the write action until the end of the block. Our proposed solution is to add to each row a D flip-flop clocked by the Tag bit. 
Prior to processing a block, write enable signals are discharged. Hence, as we traverse the LUT to process the passes of the block, a match for a row will have its Tag bit toggle from `0' to `1', setting the write enable signal at the output of the flip-flop to a `1'. At the end of each block, the rows for which the flip-flop outputs are ‘1’ are overwritten. This ensures that all rows that have matches within the block will be overwritten together by the same output. The flip-flop is reset after every block. Timewise, the flip-flop’s toggling time cost can be hidden, whereas storage wise, one extra flip-flop per row is needed.


\section{Results and Analysis}
\label{sec:results}
In this section, we study the proposed ternary adder implementations (non-blocked and blocked) in terms of energy, delay and area, and we compare them against the binary AP adder and other ternary adder implementations. First, we study the characteristics of the ``3T3R'' cell. For our experimental results, we rely on 45nm predictive technology model \cite{asu} for our simulations. The transistor threshold voltage is $V_t=0.4V$, and $V_{DD}=0.8V$. 

\subsection{Design Space Exploration: QCAM Cell Dynamic Range and Energy Analysis}
In our ternary AP adder design, a 1-trit addition involves the comparison of the key-mask pair with three stored triplets of memristor states $(M_1,\, M_2,\, M_3)$, where each triplet corresponds to one of the trits $A_i$, $B_i$ and $C_{in}$. The outcome of the comparison can result in a: full match (fm), one mismatch (1mm), two mismatches (2mm) or three mismatches (3mm).

\begin{figure}
 \centering
\includegraphics[width=0.75\linewidth]
{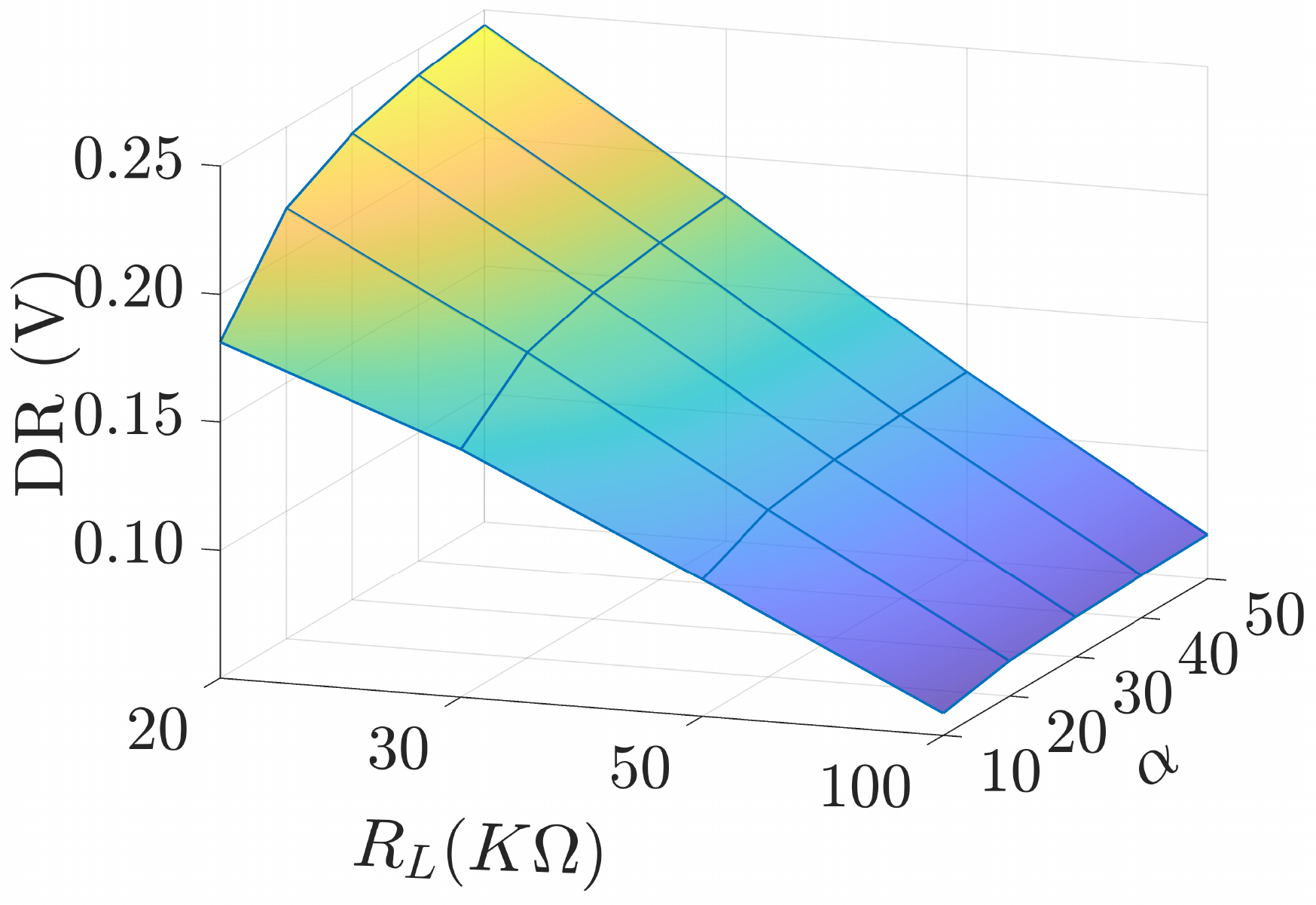}
\caption{Dynamic range for the ``3T3R'' cell for 20-trit addition as a function of $R_L$ and $\alpha$.}
\label{fig:3T3R_DR}
\vspace{-0.15in}
\end{figure}

\begin{figure}
 \centering
\includegraphics[width=0.75\linewidth]
{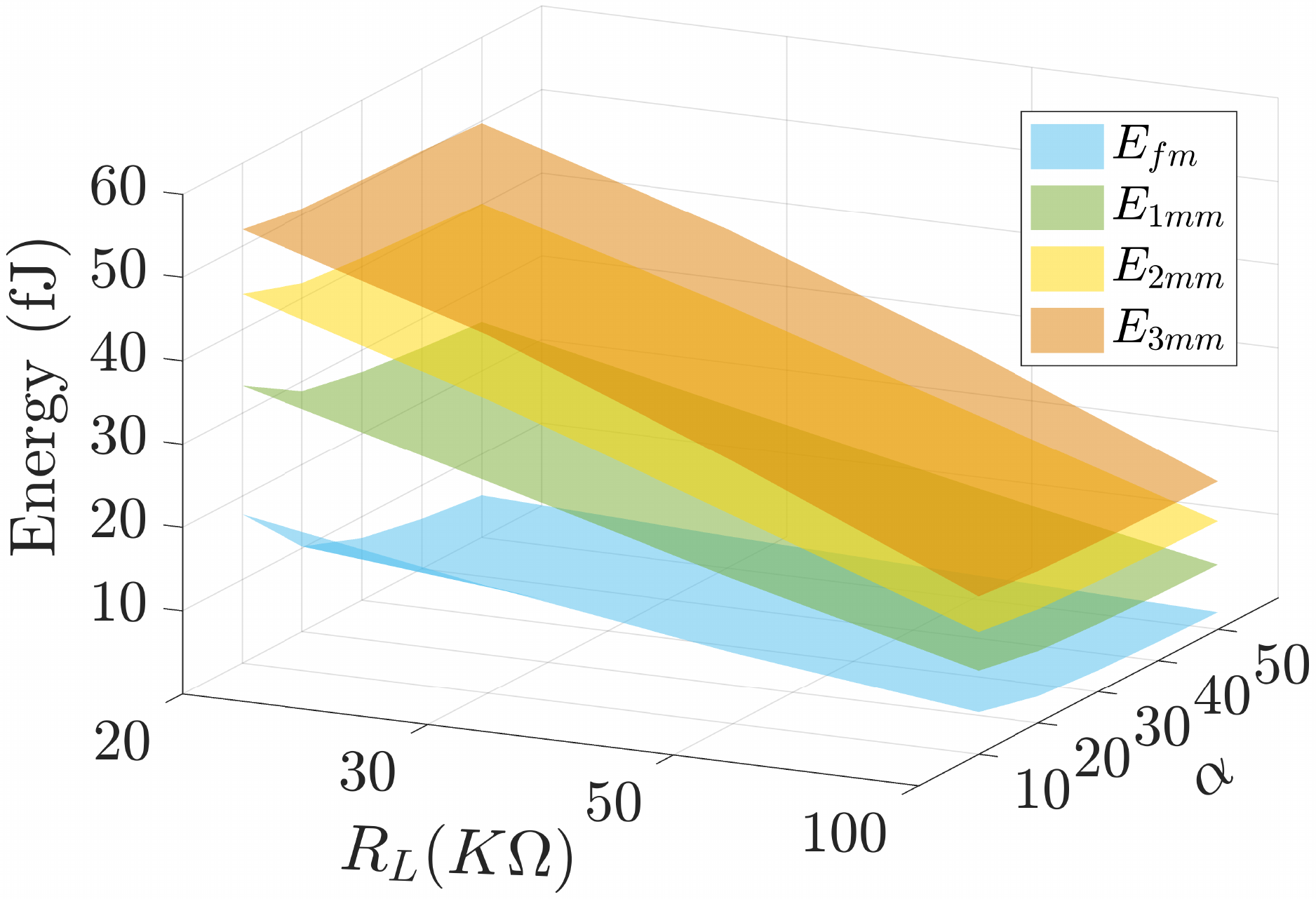}
 \caption{Compare energy for the ``3T3R'' cell for 20-trit addition as a function of $R_L$ and $\alpha$.}
 \label{fig:3T3R_energy}
 \vspace{-0.15in}
\end{figure}

For purposes of the analysis of the ``3T3R'' cell, we define the dynamic range (DR) as the maximum voltage difference between the fm and the closest mismatch case which is 1mm \cite{bahloul2017design,rakka2020design}, measured after 1ns of evaluation time of the $(S_h,\, S_x,\, S_L)$ signal triplet as indicated below. 
\begin{equation}
DR=V_{fm}-V_{1mm}
\end{equation}
where, $V_{fm}$ and $V_{1mm}$ represent the ML voltages for the fm and 1mm states, respectively. Typically, for accurate sensing of the comparison outcome, we aim for a high dynamic range. However, a high DR comes at the expense of increased compare energy consumption. 
To further assess this, we define for purposes of the add operation the compare energies $E_{fm}$, $E_{1mm}$ , $E_{2mm}$, $E_{3mm}$ corresponding to the fm, 1mm, 2mm and 3mm states, respectively. We rely on HSPICE simulations to study the dynamic range and compare energies for the ``3T3R'' cell in the context of the LUT-based ternary adder. We assess these metrics for the following design space parameter combinations.
Without loss of generality, we set the total number of cells per row $N=41$ to enable 20-trit addition, where each of the $A$ and $B$ vectors have 20 cells per vector, and we have one extra cell for the $C_{in}$ trit. We also sweep $R_L \in \{20, 30, 50, 100\} K\Omega$, and set $R_H=\alpha*R_L$ where $\alpha \in \{10, 20, 30, 40, 50\}$. A capacitive load $C_L=100fF$ is used for the comparator to properly latch $V_{ML}$ and distinguish between the fm and 1mm states due to fast discharge. 
Fig. \ref{fig:3T3R_DR} presents the corresponding dynamic range values for the ``3T3R'' cell for 20-trit addition as a function of $R_L$ and $\alpha$. The maximum, thus, best dynamic range is observed for lowest $R_L$ values. For example, $DR \approx 240mV$ when $R_L=20K \Omega$ and $\alpha=50$. The compare energy for the ``3T3R'' cell for 20-trit addition as a function of $R_L$ and $\alpha$ is plotted in Fig. \ref{fig:3T3R_energy}. For the same $R_L=20K \Omega$, the lowest energy is obtained at the highest $\alpha=50$. In fact, for $R_L=20K \Omega$, when $\alpha$ increases from 10 to 50, $E_{fm}$ drops by $71.61\%$, $E_{1mm}$ drops by $22.27\%$, $E_{2mm}$ drops by $9.45\%$ and $E_{3mm}$ drops by $4.37\%$. 
As such, for the remaining experiments, we adopt the memristor values $(R_L, R_H) = (20K\Omega, 1M\Omega)$ which provides the best dynamic range with the corresponding lowest compare energy consumption for this $R_L$ value. 

\subsection{Evaluation Against Binary System}
As previously mentioned, the optimal number system is found to be $e$ which lies in between binary and ternary system. Hence, we evaluate the TAP performance against the Binary AP. Herein, we compare the ternary LUT-based addition to the respective binary LUT-based addition in terms of energy and area. For this, we employ the non-blocked approach since these metrics are common to both the blocked and non-blocked approaches. The following section analyzes the delay of the different proposed approaches. 

Thus, in our experiments, we study the average energy for $p$-trit addition in comparison to the equivalent $q$-bit addition for different $p$ and $q$ values, where $p \in \{5t,10t,20t,32t,40t,80t\}$ and $q \in \{8b,16b,32b,51b, 64b, 128b\}$, respectively. For example, we compare $p=20t$ representing 20-trit addition to the equivalent $q=32b$ representing 32-bit addition. 

We rely on HSPICE to characterize the compare energy for 1-bit (1-trit) addition in the context of LUT-based binary (ternary) adder for the 2T2R \cite{2T2R_fouda} (3T3R) cell with $R_{L} = 20K\Omega$ and $R_{H}=1M \Omega$. We adopt $C_L=100fF$ to correctly latch $V_{ML}$, and the number of cells per row is equal to $2q+1$ ($2p+1$). We set the evaluate time to 1ns for which we observe a DR approximately equal to 200mV for the different simulations, allowing for good differentiation between the match and mismatch cases. The precharge time is also set to 1ns. 

We developed a functional simulator using MATLAB to obtain the average for the compare energy and write energy for both the ternary and binary LUT addition, relying on a total of 10,000 $p$-trit and $q$-bit additions. The functional simulator estimates the number of set/reset operations taking into consideration whether we are writing only one, two, or all three $(A,\, S,\, C_{out})$ cells based on the different LUTs and number of sets/resets required per cell (see Table \ref{tab:set_reset}). We assume the memristor write energy per set or reset operation to be on average around 1nJ as was stated in \cite{write_energy} for different programming and initial memristor conditions. The functional simulator also utilizes the 1-bit and 1-trit compare energy values obtained using HSPICE to estimate the $q$-bit ($p$-trit) compare energy based on the different match/mismatch combinations. 
For purposes of area comparison, we rely on the number of cells per row for the $q$-bit ($p$-trit) addition assuming that the ``2T2R'' cell area is 0.67x the area of one ``3T3R'' cell. Results are indicated in Table \ref{tab:energy_area}. Overall, compared to the LUT-based binary addition, the LUT-based ternary addition results in about 12.6\% reduction in the total number of sets/resets needed, 12.25\% reduction in total energy and 6.2\% area reduction. 

\begin{table*}[]
\centering
\caption{Energy and area comparison of the ternary AP adder with the binary AP adder \cite{binary_AP}.}
\label{tab:energy_area}
\begin{tabular}{ll|cc|cc|cc|cc|cc|cc|}
\cline{3-14}
       &      & \textbf{8b} & \textbf{5t} & \textbf{16b} & \textbf{10t} & \textbf{32b} & \textbf{20t} & \textbf{51b} & \textbf{32t} & \textbf{64b} & \textbf{40t} & \textbf{128b} & \textbf{80t} \\ \hline
\multicolumn{1}{|l|}{\multirow{4}{*}{Energy}} & \#Set = \#Reset & 5.99 & 5.22 & 11.99 & 10.53 & 24.04 & 21.02 & 38.24 & 33.67 & 47.98 & 42.17 & 95.98 & 84.54 \\ \cline{2-14} 
\multicolumn{1}{|l|}{} & Write energy (nJ)  & 11.99 & 10.44 & 23.99 & 21.06 & 48.07 & 42.04 & 76.48 & 67.35 & 95.96 & 84.33 & 192.0 & 169.1 \\ \cline{2-14} 
\multicolumn{1}{|l|}{} & Compare energy (pJ) & 0.94 & 3.99 & 1.91 & 8.06 & 3.90 & 16.4 & 6.36 & 26.84 & 8.11 & 34.0 & 17.5 & 72.58 \\ \cline{2-14} 
\multicolumn{1}{|l|}{} & Total energy (nJ)  & 11.99 & 10.44 & 23.99 & 21.07 & 48.07 & 42.06 & 76.49 & 67.38 & 95.97 & 84.36 & 192.02 & 169.17 \\ \hline
\multicolumn{2}{|l|}{Normalized Area} & 16x & 15x & 32x & 30x & 64x & 60x & 102x & 96x & 128x & 120x & 256x & 240x \\ \hline
\end{tabular}
\vspace{-0.1in}
\end{table*}

\subsection{Performance Comparison to Other Ternary Adder Designs}
In this section, we compare the proposed ternary LUT-adder approaches against other ternary adder implementations.
Particularly, we compare the total energy consumed by the proposed ternary AP (TAP) adder implementations against hybrid CNTFET and memeristor-based implementations of the carry-ripple adder (CRA), carry-skip adder (CSA) and carry-lookahead adder (CLA) \cite{other_adders}. We also conduct a delay analysis for the TAP blocked and non-blocked approaches in comparison to the LUT-based binary adder and the CLA. 

\begin{figure}
\centering
\includegraphics[width=0.85\linewidth]{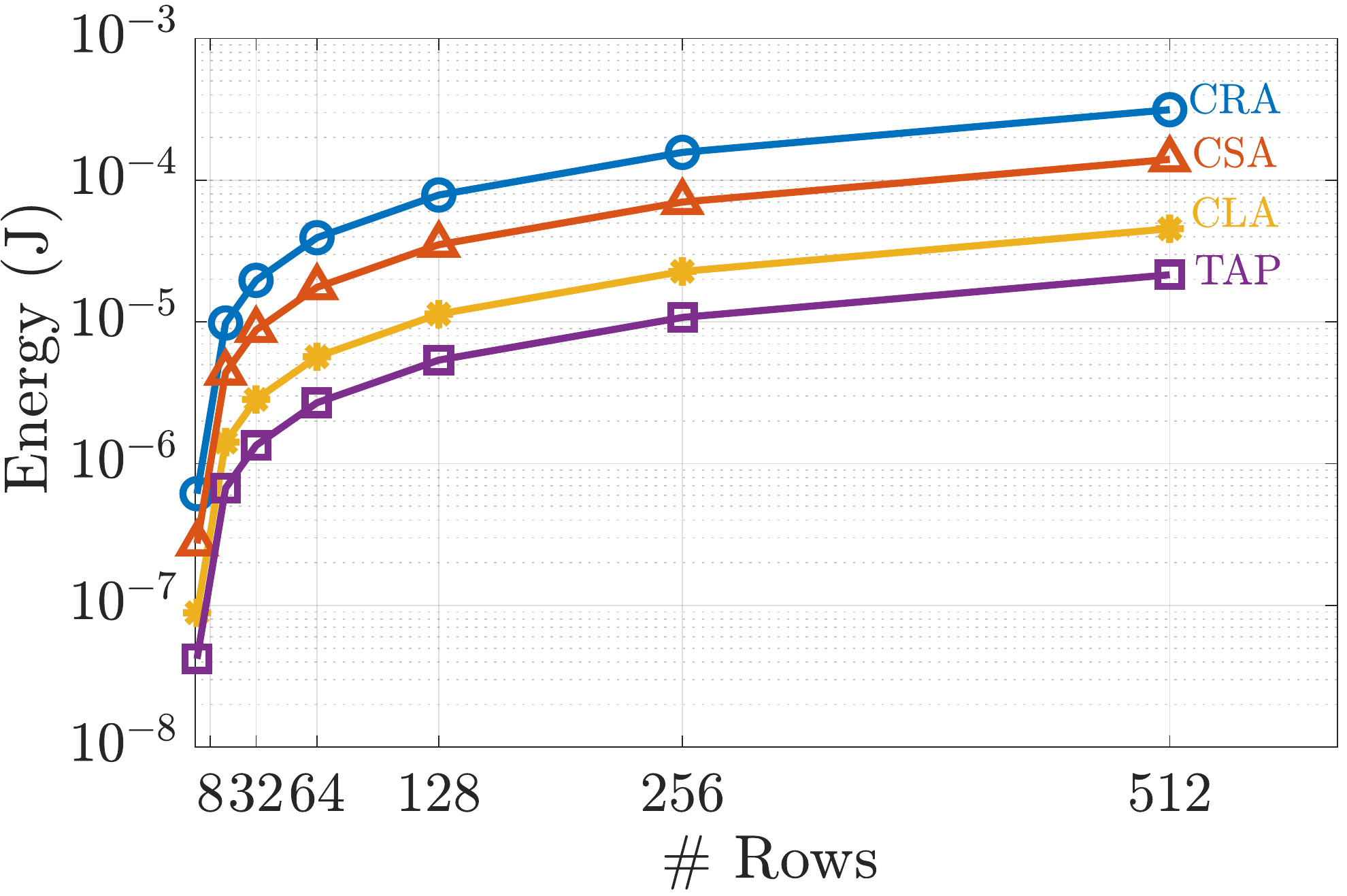}
\caption{Energy comparison of the TAP versus other ternary adders \cite{other_adders} for set/reset energy of {1nJ}.}
\label{fig:energy_comparisons}
\vspace{-0.1in}
\end{figure}

Our comparison is based on extrapolating the authors' 4-bit adder's power and delay simulations to reflect energy consumption and delay values for 20-trit addition at $V_{DD}=0.8V$. 
For our adder implementation, the consumed energy does not differ between the non-blocked and blocked approaches. We thus rely on Table \ref{tab:energy_area} to obtain the total energy for our TAP implementation.
Fig. \ref{fig:energy_comparisons} presents the energy for the different ternary adder implementations as function of the number of rows (\#Rows, i.e., number of parallel additions). TAP consumes about 52.64\% less energy than the CLA, which in turn demonstrated lower energy consumption compared to the CSA and CRA. We note that for all adder implementations, the energy grows linearly with the number of add operations.



\begin{figure}
 \centering
\includegraphics[width=0.75\linewidth]
{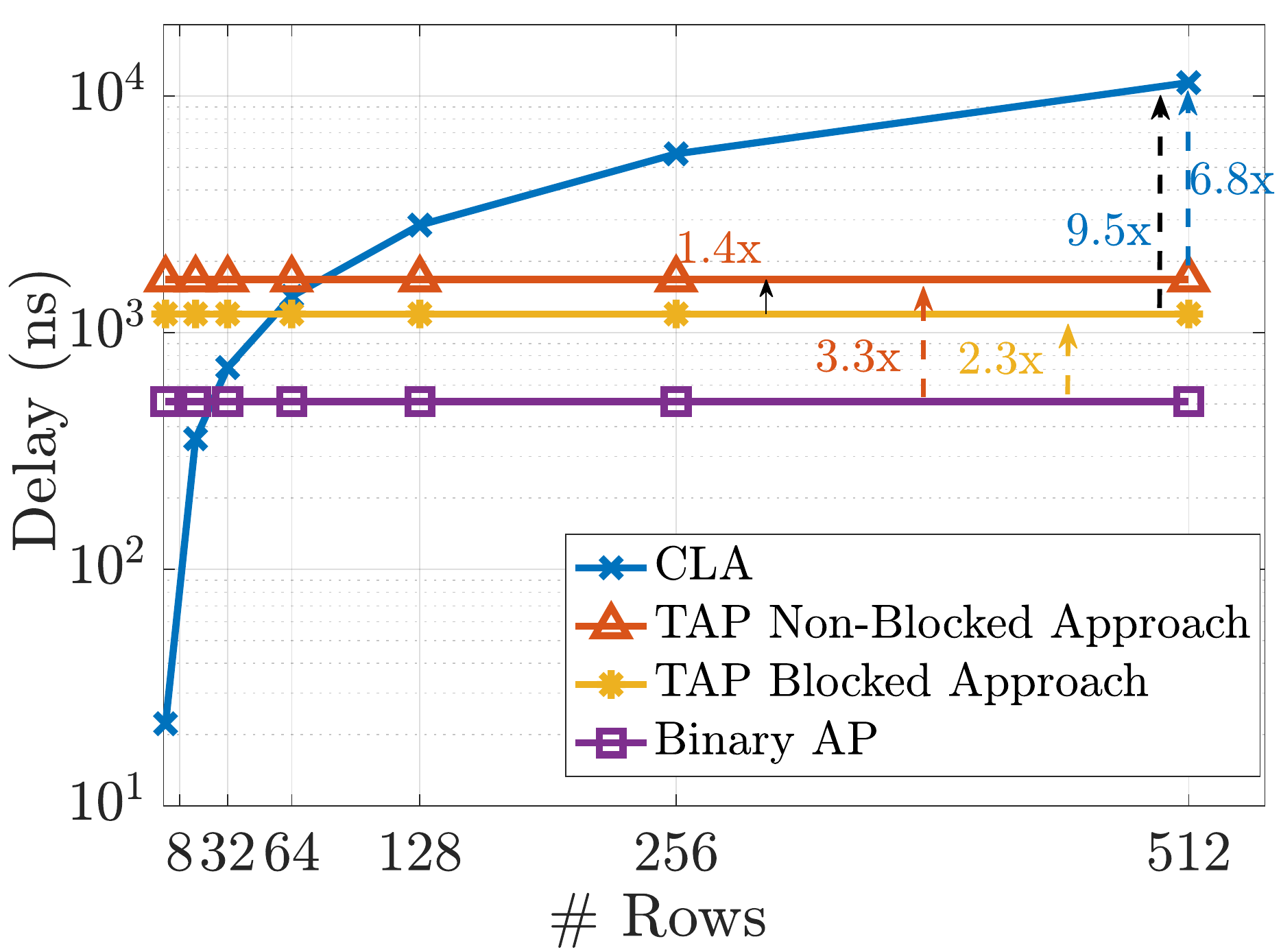}
 \caption{Delay comparison for the blocked and non-blocked TAP implementations with the CLA \cite{other_adders} and the binary AP adder \cite{binary_AP}.}
 \label{fig:delay_comparisons}
 \vspace{-0.1in}
\end{figure}

We define the delay as the number of clock cycles needed to concurrently compare and write multiple rows within the data array. While in the non-blocked approach every compare is followed by a write action, the blocked approach delays the write action until the end of the sequence of block compares, thus improving the overall delay of the adder. Note that, irrespective of whether a match occurs or not, we account for the write cycle. 
Fig. \ref{fig:delay_comparisons} shows the delay for the LUT ternary adder using the non-blocked and blocked approaches, along with the CLA and the binary AP adder as a function of the number of rows (\#Rows). Compared to the CLA, the non-blocked (blocked) TAP demonstrates lower runtime starting when the number of $p$-trit add operations (i.e., \#Rows) exceeds 64 (32). At 512 rows, the non-blocked and blocked TAP approaches demonstrate a 6.8x and 9.5x reduction in delay compared to the CLA, respectively. The blocked approach further shows a 1.4x reduction in delay compared to the non-blocked approach for all \#Rows. These results assume a traditional precharge cycle similar to Fig. \ref{fig:timing_diagram}.

In an optimized implementation where the precharge is embedded within the write cycle, the TAP adder is 9x smaller than the CLA and the blocked TAP approach introduces around 1.2x improvement compared to the non-blocked TAP approach. This is due to the need for precharge post evaluate for the compare cycles that are not followed by a write action. 

Finally, it is worth noting that the binary AP adder demonstrates the lowest delay at 2.3x savings compared to the ternary TAP in lieu of increased area and energy.





\section{Conclusion}
In this paper, we proposed a novel multi-valued associative processor with illustrative example on ternary-radix. In addition, we proposed efficient LUT-based ternary full adder methodology in the context of AP. The AP implementation relies on a novel quaternary CAM ``3T3R'' cell. Novel algorithms are used to build the ternary adder LUT following two approaches: a first non-blocked approach that formalizes the intuition behind LUT pass ordering and a second blocked approach that targets latency reduction in terms of capitalizing on common write action cycles. The efficiency of the proposed approaches is proven by a formal simulator built using MATLAB in which we incorporate HSPICE simulations. Results show that the ternary AP has lower energy and area compared to the binary AP, albeit higher delay. Moreover, compared to other hybrid CNTFET and memristor implementations of the ternary adder, the proposed ternary AP adder has lower energy and delay. Furthermore, the results demonstrate performance efficiency for the blocked AP approach. In order to improve the performance of MvAP, multi-valued CAM cell would need to optimized with less number devices and with more efficient write techniques. 




\begin{IEEEbiography}[{\includegraphics[width=1in,height=1.25in]{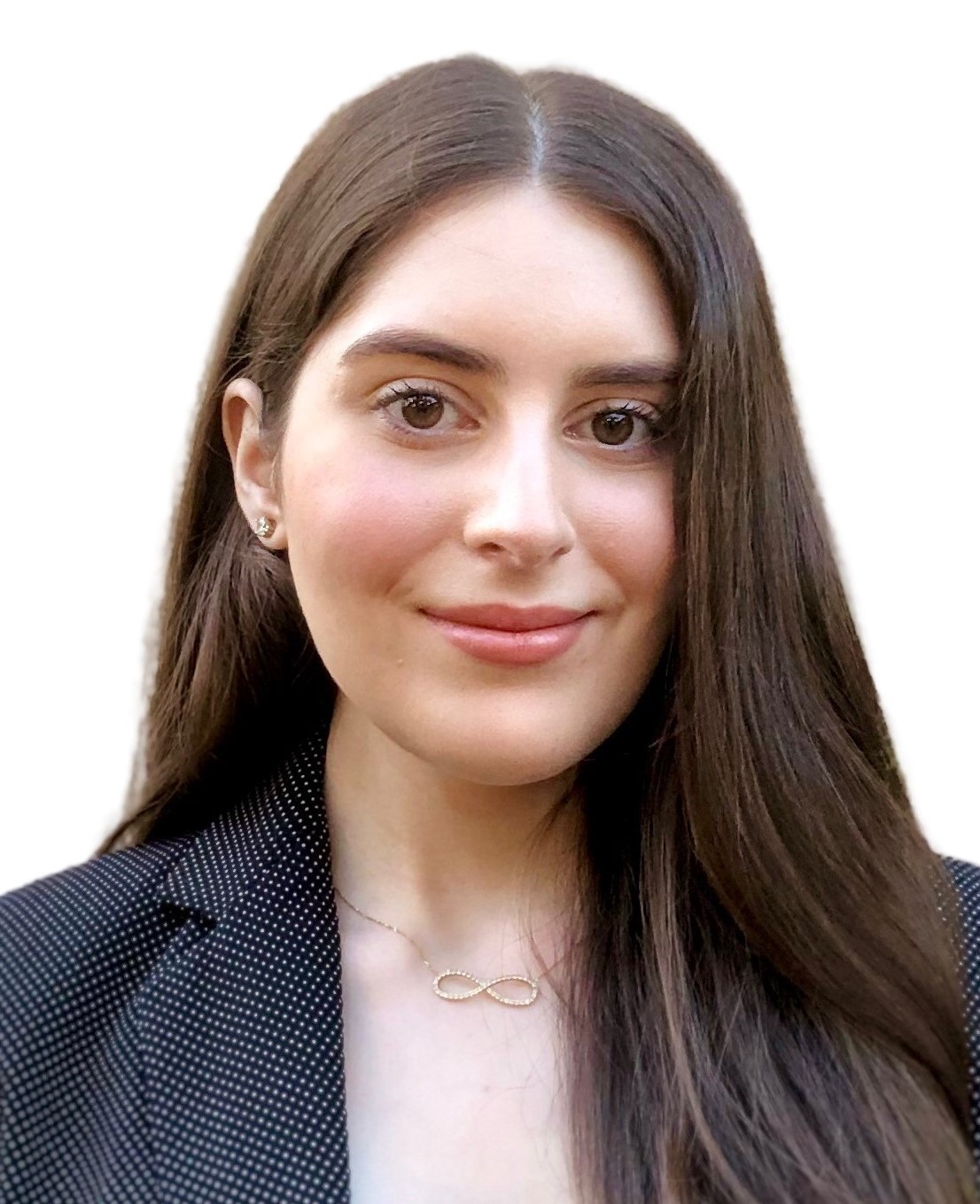}}]{Mira Hout} (SM 2017) received a B.Eng. degree in Computer and Communications Engineering (with High Distinction) from the Maroun Semaan Faculty of Engineering and Architecture, American University of Beirut, Lebanon, in 2021. During her senior year, she worked as a research assistant where she focused on topics of associative processors for purposes of ternary logic ReRAM-based arithmetic circuits design. She also worked as a teaching assistant in the Department of Electrical and Computer Engineering at AUB for electronics courses. She is currently a research intern at King Abdullah University of Science and Technology with the Department of Electrical Engineering and Computer Science in the scope of an ongoing collaboration between AUB and KAUST. Her work revolves around studying multi-valued CAM-based associative processors in the context of in-memory arithmetic and logic compute. Her research interests include multi-valued logic, emerging ReRam-based methodologies for arithmetic circuit designs and in-memory computing.
\end{IEEEbiography}

\begin{IEEEbiography}[{\includegraphics[width=1in,height=1.25in]{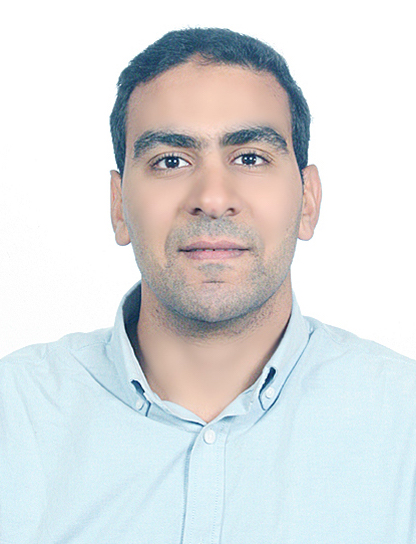}}]{Mohammed E. Fouda} received a B.Sc. degree (Hons.) in Electronics and Communications Engineering and an M.Sc. degree in Engineering Mathematics from the Faculty of Engineering, Cairo University, Cairo, Egypt, in 2011 and 2014, respectively. Fouda received his Ph.D. degree from the University of California, Irvine, USA in 2020. Currently, He works as an assistant researcher at University of California, Irvine. Fouda has published more than 100 peer-reviewed Journal and conference papers, one Springer book, and three book chapters. His H-index is 21, with more than 1600 citations. His research interests include analog AI hardware, neuromorphic circuits and systems, brain-inspired computing, memristive circuit theory, fractional circuits, and systems and analog circuits. He serves as a peer-reviewer for many prestigious journals and conferences. He also serves as a review editor in Frontier of Electronics and the International Journal of Circuit theory and applications, in addition to serving as a technical program committee member in many conferences. He was the recipient of the best paper award in ICM years 2013 and 2020 and the Broadcom foundation fellowship for 2016--2017.
\end{IEEEbiography}

\begin{IEEEbiography}[{\includegraphics[width=1in,height=1.1in]{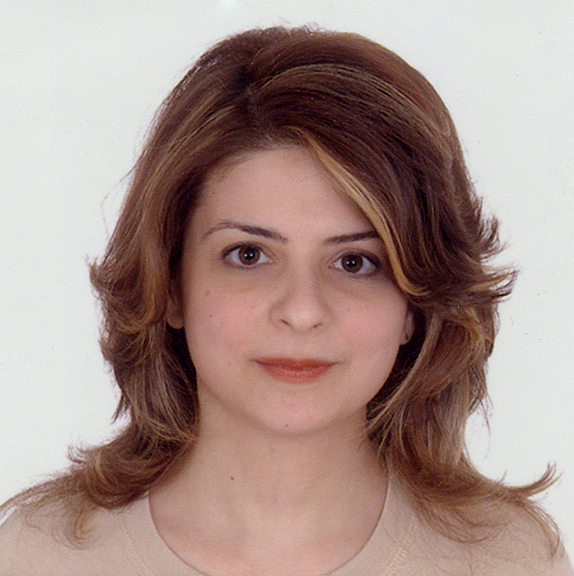}}]{Rouwaida Kanj} received the M.S. and Ph.D. degrees in electrical engineering from the University of Illinois Urbana-Champaign in 2000 and 2004, respectively. She is currently a tenured Associate Professor at the American University of Beirut. From 2004-2012 she worked as a research staff member at IBM Austin Research Labs. Her research work focuses on advanced algorithmic research and development and smart analytics methodologies for Design for Manufacturability Reliability and Yield with emphasis on statistical analysis, optimization and rare fail event estimation for microprocessor memory designs along with machine learning applications for Very Large Scale Integration. More recently she is also involved in memristor-based memory design and reliability and the design of reliable circuits and systems for healthcare and in-memory compute. This is in addition to her earlier work on noise modeling and characterization of CMOS designs. Dr. Kanj was a recipient of three IBM Ph.D. Fellowships, is the author of more than 81 technical papers, 36 issued US patents and several pending patents. She received an outstanding technical achievement award and 6 Invention Plateau awards from IBM. She received the prestigious IEEE/ACM WILLIAM J. MCCALLA ICCAD best paper award in 2009, and two IEEE ISQED best paper awards in 2006 and 2014, and the IEEE ICM best paper award in 2020. In 2018 her work on statistical yield analysis methodology was nominated for the ACM/IEEE Richard Newton Award. She is currently a senior member of IEEE and serves or chairs on the technical program committees of several prestigious IEEE conferences. She also serves as a peer-reviewer for many prestigious journals.  
\end{IEEEbiography}

\begin{IEEEbiography}[{\includegraphics[width=1in,height=1.25in,clip,keepaspectratio]{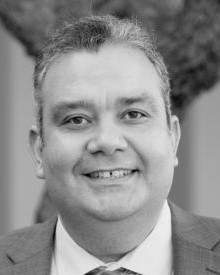}}]{Ahmed E. Eltawil}(S’97–M’03–SM’14) received a Doctorate degree from the University of California, Los Angeles, in 2003 and M.Sc. and B.Sc. degrees (with honors) from Cairo University, Giza, Egypt, in 1999 and 1997, respectively. Since 2019, he has been a Professor at the Computer, Electrical, and Mathematical Science and Engineering Division at the King Abdullah University of Science and Technology, Thuwal, KSA. Since 2005, he has been with the Electrical Engineering and Computer Science Department at the University of California, Irvine. His research interests are in the general area of low-power digital circuit and signal processing architectures, with an emphasis on mobile systems. Dr. Eltawil has been on the technical program committees and steering committees for numerous workshops, symposia, and conferences. He has received several awards, as well as distinguished grants, including the NSF CAREER grant supporting his research into low-power systems.
\end{IEEEbiography}

\end{document}